\documentclass[mathpazo]{cicp}

\usepackage{graphicx}
\usepackage{bm}
\usepackage{braket}

\begin{document}
\title{Inertial Frame Independent Forcing for Discrete Velocity Boltzmann Equation: Implications for Filtered Turbulence Simulation}


\author[K.N. Premnath]{Kannan N. Premnath\affil{1}\comma\corrauth, and Sanjoy Banerjee\affil{2}}

\address{\affilnum{1}\ Department of Mechanical Engineering,
          University of Wyoming,
          Laramie, WY 82071, U.S.A. \\
         \affilnum{2}\ Department of Chemical Engineering, City College of New York, City University of New York, New York, NY 10031, U.S.A.}

\emails{{\tt knandhap@uwyo.edu} (K.N.Premnath)}

%


\begin{abstract}
We present a systematic derivation of a model based on the central moment lattice Boltzmann equation that rigorously maintains Galilean invariance of forces to simulate inertial frame independent flow fields. In this regard, the central moments, i.e. moments shifted by the local fluid velocity, of the discrete source terms of the lattice Boltzmann equation are obtained by matching those of the
continuous full Boltzmann equation of various orders. This results in an exact hierarchical identity between the central
moments of the source terms of a given order and the components of the central moments of the distribution functions and sources of
lower orders. The corresponding source terms in velocity space are then obtained from an exact inverse transformation due to a suitable choice of orthogonal basis for moments. Furthermore, such a central moment based kinetic model is further extended by incorporating reduced compressibility effects to represent incompressible flow. Moreover, the description and simulation of fluid turbulence for full or any subset of scales or their averaged behavior should remain independent of any inertial frame of reference. Thus, based on the above formulation, a new approach in lattice Boltzmann framework to incorporate turbulence models for simulation of Galilean invariant statistical averaged or filtered turbulent fluid motion is discussed.
\end{abstract}


\pac{05.20.Dd,47.27.-i,47.27.E-}

\maketitle

\section{Introduction} \label{section_introduction}
Minimal kinetic models for the Boltzmann equation, i.e. lattice Boltzmann equation formulations, are evolving towards
as alternative physically-inspired computational techniques for various fluid mechanics and other problems.
Originally developed as an improved variant of the lattice gas automata~\cite{frisch86} to eliminate statistical noise~\cite{mcnamara88}, the lattice Boltzmann method (LBM) has undergone a series of major refinements, in terms of its underlying physical models as well as numerical solution schemes for various applications over the last two decades~\cite{benzi92,chen98,succi01,yu03}. In particular, its rigorous connection to the kinetic theory~\cite{abe97,he97,shan98} has resulted in a number of recent developments, including models that are more physically consistent for multiphase~\cite{luo00,he02} and multicomponent flows~\cite{asinari08}, models for non-equilibrium phenomena beyond the Navier-Stokes-Fourier representation~\cite{shan06} and an asymptotic analysis approach to establish consistency of the LBM from a numerical point of view~\cite{junk05}.

The stream-and-collide procedure of the LBM can be considered as a dramatically simplified discrete representation of
the continuous Boltzmann equation. Here, the streaming step represents the advection of the distribution of particle
populations along discrete directions, which are designed from symmetry considerations, between successive collisions.
Much of the physical effects being modeled are represented in terms of the collision step, which also significantly influences
the numerical stability of the LBM. Most of the major developments until recently were associated with the single-relaxation-time (SRT) model~\cite{qian92,chen92} based on the BGK approximation~\cite{bhatnagar54}, and enjoys its popularity owing, mainly, to its simplicity. However, it is prone to numerical instability. Moreover, it is inadequate in its representation of certain physical aspects, such as independently adjustable transport properties of thermal transport and viscoelastic phenomena.

These limitations have been significantly addressed in the multiple-relaxation-time (MRT) collision model~\cite{dhumieres92}.
This, in a sense, represents a simplified form of the relaxation LBM proposed earlier~\cite{higuera89a,higuera89b}, with an important characteristic difference in that the collision process is carried out in moment space~\cite{grad49} instead of in the usual velocity space. By separating the relaxation time scales of different moments, obtained by using a linear Fourier stability analysis, its numerical stability can be significantly improved~\cite{lallemand00,dhumieres02}. Furthermore, it has resulted in significant advantages over the SRT-LBM for computation of various classes of fluid flow problems, including multiphase systems~\cite{premnath06,premnath05a,premnath05b}, turbulent flows~\cite{premnath09a,premnath09b} and magnetohydrodynamics~\cite{pattison08}. It may be noted that recently a different form of MRT model based on the orthogonal Hermite polynomial projections of the distribution functions, which is independent of any underlying lattice structure, allowing representation of higher order non-equilibrium effects has been proposed~\cite{shan07}.

The stabilization of the LBM using a single relaxation time has been addressed from a different perspective by enforcing the H-theorem locally in the collision step~\cite{karlin99,boghosian01,ansumali02,succi02}. By using the attractors of the distribution function
based on the minimization of a Lyapunov-type functional, non-linear stability of the LBM is achieved in this Entropic LBM.
This approach has recently been significantly extended to incorporate multiple relaxation times with efficient implementation
strategies~\cite{asinari09,asinari10}. Furthermore, systematic procedures for different types of higher-order LBM have been
developed~\cite{chikatamarla09,dubois09,karlin10}. An important element is the construction of higher-order lattices based on
symmetry considerations which have been analyzed using group theory~\cite{rubinstein08,chen08}. Further progress, from a numerical
aspect, is that based on the consistency analysis~\cite{junk05} and a notion of structural stability~\cite{banda06,junk09a} (shown
related to the Onsager-like relation in non-equilibrium thermodynamics~\cite{yong09}), convergence of the LBM to
the Navier-Stokes equations has rigorously been shown~\cite{junk09b}.

On the other hand, it is important to clearly understand in what sense the lattice Boltzmann equation (LBE), which is generally considered
as a mesoscopic approach, inherits or maintains the various physical invariance properties of the continuous full Boltzmann equation (which
it represents as a much simplified model) and the Navier-Stokes equations (which it represents numerically). Careful considerations of these aspects play an important role in ensuring the general applicability of the approach for various, especially challenging, problems. In this regard, and to put the present work in perspective, it should be noted that the continuum mechanics description as well as the microscopic statistical (continuous Boltzmann) description of fluid motion generally satisfy a larger invariance group, with inertial frame invariance being just an important special case. The most general form among these is the so-called the principal of material frame indifference, also known as the objectivity principle~\cite{truesdell04}. According to this, the constitutive equations should have the same forms in \emph{all} frames of reference, whether inertial or not. While this is considered as an important axiom based on which the continuum mechanics is formulated~\cite{truesdell04}, its role from continuous kinetic theory point of view was the subject of considerable analysis for sometime~\cite{muller72,wang75,speziale81,woods83,speziale86,speziale87}.

The following are the main outcomes of these studies: the continuous full Boltzmann equation (i) is material frame indifferent in a \emph{strong approximate sense}, when there is a large scale separation between the collision times and the macroscopic flow times~\cite{speziale81,speziale86,speziale87} (thus providing a strong support to the axiomatic principle generally used in the continuum description), (ii) satisfies both the inertial frame or Galilean invariance as well as the extended Galilean invariance (i.e. invariance under arbitrary translational accelerations of the reference frame) \emph{exactly}~\cite{speziale87}. Furthermore, it was shown that the standard procedures (e.g. Chapman-Enskog expansion, Maxwellian iteration) lead to frame dependent higher order contributions for the constitutive equations in non-inertial frames in the continuum limit, while the continuous kinetic theory itself can be frame independent~\cite{speziale81}. Careful considerations of these principles could guide in the development of more generally applicable models and numerical schemes for complex problems. For example, material frame indifference (point (i)) is generally used as an important constraint for the constitutive equations for complex fluids (e.g. beyond Newtonian constitutive description such as polymeric fluids) and in the development of turbulence models in continuum mechanics. As mentioned above, this property is satisfied in a strong approximate sense by the continuous Boltzmann equation, but not necessarily by the tools that relate the microscopic and macroscopic descriptions(point iii). This aspects are pertinent in the construction of complex models from the continuous Boltzmann equation (e.g.~\cite{wagner01,degond02}). In this work, however, we limit our discussion to an association of the properties mentioned in a part of the point (ii) for the LBE, i.e. for the exact invariance group -- the Galilean or inertial frame invariance.

In this regard, as a discrete approximation to the continuous full Boltzmann equation, the development of the LBE consists of
simplifications at different levels. Thus, its various elements should be analyzed carefully to ascertain and quantify as to how
well it satisfies Galilean invariance. First, in contrast to continuous kinetic theory, due to the choice of finite lattice velocity sets and associated symmetries, it introduces linear dependencies of higher order moments with those of lower order moments that are supported by the lattice set~\cite{rubinstein08}. Such degeneracies can in turn lead to negative dependence of viscosity on fluid velocity. It generally causes the Galilean invariance to be broken by the presence of terms that are cubic in velocity for the standard lattice configurations (with symmetries of square in 2D and cube in 3D) and also leads to numerical instability, especially at higher Mach numbers. This issue can be alleviated by the use of extended lattice velocity sets, which then relegates the degeneracies among moments to even higher orders. Second, the collision step including the forcing terms of the LBE should be carefully constructed in such a way that they recover correct physics which is inertial frame independent, i.e. the Navier-Stokes equations. Here, the use of independent set of \emph{central} moments for a chosen lattice provides a natural approach to maintain Galilean invariance that can be constructed by invoking elements directly from kinetic theory. This is the main goal of the present work (see below). A rational means to more efficiently account for both the above aspects is discussed in the last section of this paper. And, third, the streaming step of the LBE is generally constructed as a discrete Lagrangian process. In the standard implementation, this couples the particle velocity and configurations spaces, which in turn, constrains the numerical accuracy of the LBE in the representation of the Navier-Stokes equations. As a result, the Galilean invariance of the LBE is limited by its overall numerical accuracy. However, it is known that such coupling between physical and lattice symmetries is not necessary in the discretization of the streaming operator. In fact, it can be discretized using classical schemes such as finite-difference or finite-element methods that alleviate this issue~\cite{cao97,sofonea03,lee03}. Specifically, exploiting higher order discretization and time integration schemes (e.g.~\cite{min11}) for the streaming operator could further improve the order of accuracy (and hence the Galilean invariance) of the LBE. Furthermore, the use of implicit schemes could enhance the computational efficiency in this regard.

Focusing on the second aspect mentioned above, a different type of collision operator and forcing can be devised that can maintain Galilean invariance for a chosen lattice velocity set and a discretization scheme for the streaming step. Specifically, central moments are relaxed in a moving frame of reference during collision step~\cite{geier06}, originally proposed to improve numerical stability, but emphasized here for its better physical coherence. The use of central moments, which are obtained by shifting the particle velocity with the local fluid velocity~\cite{struchtrup05}, rigorously enforces Galilean invariance. In particular, while other previous approaches are generally Galilean invariant for up to second-order moments, the central moment based approach provides a higher order frame invariance as permitted by the discrete lattice velocity set. This approach was examined based on the concept of generalized local equilibrium~\cite{asinari08new}. In addition, to further improve physical coherence, the attractors for the higher order central moments were constructed as products of the lower order central moments, leading to the factorized central moment method~\cite{geier09}. Recently, a new approach to incorporate source terms using central moments in the LBM that are Galilean invariant by construction, which are important for computation of various physical problems, was developed~\cite{premnath09d}. The consistency of this technique to the Navier-Stokes equations was shown by means of the Chapman-Enskog analysis~\cite{chapman64} and its numerical accuracy was established. Furthermore, the method was also extended in three-dimensions for various lattice velocity sets and validated for a class of canonical problems~\cite{premnath11}. As clarified in~\cite{premnath09d,premnath11}, numerical stability of the central moment approach can be enhanced, when it is executed in a multiple relaxation time formulation, similar to the standard or raw moment based approaches. Interestingly, it has been shown recently that when
some classical schemes for flow simulation are made to satisfy Galilean invariance more rigorously, they led to more robust implementations (e.g.~\cite{scovazzi07a,scovazzi07b}).

Turbulence remains as among the most challenging classes of flows for which considerable effort has been focused on the
development of theory and applications using the LBM. Since its roots can be traced to kinetic theory, the LBM has been analyzed
for the development of turbulence models from a fundamental point of view~\cite{chen98a,ansumali04,chen04}. It has been employed for
computation of Reynolds-averaged description of turbulent flows~\cite{teixeira98,chen03}. Furthermore, it has found applications for
large eddy simulation (LES) using LBM formulations with SRT~\cite{hou96}, and MRT~\cite{yu06} with multiblock approach for efficient
implementation~\cite{premnath09a}. Recently, dynamic subgrid scale (SGS) models for LES were incorporated into the LBM framework that
resulted in reduced empiricism for description at such scales~\cite{premnath09b}. Moreover, an improved inertial-range consistent SGS
model was also proposed~\cite{dong08}. A theoretical formulation for a SGS model based on an approximate deconvolution method~\cite{stolz99} that does not rely on the common eddy-viscosity concept for application with the LBM was also devised recently~\cite{sagaut10}. Lastly, the closure modeling issues of kinetic and continuum turbulence effects were reconciled in a unified statistical/filtered description using a modified kinetic equation~\cite{girimaji07}. Effectively, this allows the use of macroscopic turbulence models involving divergence of the
Reynolds stress in the forcing term of the kinetic equation.

An important physical consideration for any description of turbulent flow is that it should be invariant for all inertial frames
of reference. In other words, for general applicability, representation of turbulence for all or any subset of its scales should be
Galilean invariant. Thus, in particular, all SGS models, and associated numerical schemes for turbulence computation, should be frame invariant. An insightful analysis of various turbulence models was carried out from this viewpoint in~\cite{speziale85}.
A method to achieve Galilean invariance by means of certain redefinition of turbulent stresses was discussed in~\cite{germano86}.
A recent review on this subject is reported in~\cite{fureby97}. Furthermore, it should be noted that concepts based on central moments have played an important role in the development of theoretical foundation of turbulence physics -- such as for statistical turbulence theory~\cite{monin07} and turbulence modeling~\cite{germano92}.

Thus, in this paper, we develop a lattice Boltzmann equation based on central moments for Galilean invariant representation of turbulent flows. Specifically, it allows frame-independent incorporation of general models for turbulent Reynolds stresses in a statistical/filter averaged formulation using LBM for turbulence simulation. Furthermore, in a general setting, it maintains the forces and stresses to be independent of any inertial frame of reference and could also improve numerical stability in computations. In~\cite{premnath09d}, we developed a forcing scheme based on a particular ansatz involving the local Maxwell distribution. Here, we develop a general forcing based on central moments by a direct examination of the continuous full Boltzmann equation itself, which unlike~\cite{premnath09d} could also self-consistently account for non-equilibrium effects in higher order terms. In this regard, the central moments of the resulting source terms of the continuous and discrete counterparts are matched successively at different orders leading to a cascaded structure. In essence, this approach can be considered as a Galilean invariant minimal discrete model for the full Boltzmann equation including forcing terms. The attractors for higher order central moments in the collision step of this computational model is based on the factorization in terms of those at lower orders by including such general forcing terms. In addition, we further develop this approach with reduced compressibility effects for improved representation of turbulent flow physics in the incompressible limit. The forcing formulation developed here for incorporating turbulence models in a statistical/filtered formulation can be extended to other problems, such as, for example, Galilean invariant representation of forces or stresses in complex fluids.

The paper is organized as follows. Section~\ref{sec:discreteparticlevelocity} briefly discusses the choice of the moment basis
employed in this paper and Sec.~\ref{sec:continuousBoltzmannequation} the continuous Boltzmann equation.
In Secs.~\ref{sec:ccentralmomentsfandfeq}~and~\ref{sec:ccentralmomentssources}, continuous forms of the central moments for the
distribution functions and its local equilibrium, and sources due to force fields, respectively, are introduced. The LBE based on
central moments with the general forcing terms is presented in Sec.~\ref{sec:cascadedLBEforcing}. Various discrete central moments
are presented in Sec.~\ref{sec:discretecentralmoments} that also specifies a matching principle to maintain Galilean invariance and the relationships among such moments are provided in Sec.~~\ref{sec:discretemomentsrelation}. Section~\ref{sec:rawmoments} describes various discrete raw moments and the derivation of the source terms in terms of the discrete particle velocity space. In Sec.~\ref{sec:cascadedcollisionforcing},we present the construction of the collision operator of the central moment based LBM. The computational procedure of this approach is provided in Sec.~\ref{sec:computationalprocedure}. The derivation is extended by considering reduced compressibility effects in Sec.~\ref{sec:cascadedcollisionreducedcompressibility}. Furthermore, Sec.~\ref{sec:factorizedcentralmomentmethod} discusses the use of attractors of the higher order central moments based
on the concept of their factorization in term of those at lower orders. A natural consequence of this overall approach is that
turbulence models can be represented for Galilean invariant filtered turbulence simulation using the LBM, which is described in Sec.~\ref{sec:Galileaninvariantturbmodels}. Finally, the summary and conclusions of this work are discussed in Sec.~\ref{sec:summary}.

\section{\label{sec:discreteparticlevelocity}Selection of Moment Basis}
\begin{figure}
\begin{center}
\includegraphics[width = 65mm]{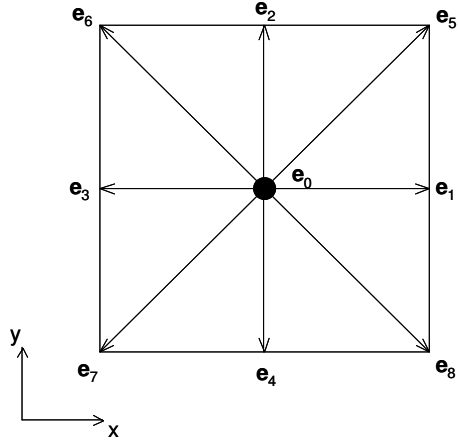}
\caption{\label{fig:d2q9} Two-dimensional, nine-velocity (D2Q9) Lattice.}
\end{center}
\end{figure}

An important element in the development of the central moment based LBM is the specification of a suitable basis for moments. In this work, to elucidate our approach, the two-dimensional, nine velocity (D2Q9) lattice model (see Fig.~\ref{fig:d2q9}) is considered, for which the moment basis used in~\cite{premnath09d} is adopted. It should, however, be noted that the procedure described henceforth is of general nature, and can be extended for other lattice models and in three dimensions. The particle velocity for this lattice model $\overrightarrow{e}_{\alpha}$ is given by
\begin{equation}
\overrightarrow{e_{\alpha}} = \left\{\begin{array}{ll}
   {(0,0)}&{\alpha=0}\\
   {(\pm 1,0),(0,\pm 1)}&{\alpha=1,\cdots,4}\\
   {(\pm 1,\pm 1)} &{\alpha=5,\cdots,8}
\end{array} \right.
\label{eq:velocityd2q9}
\end{equation}
For convenience, we employ Dirac's bra-ket notion to represent the basis vectors, and Greek and Latin subscripts for
particle velocity directions and Cartesian coordinate directions, respectively. Noting that moments in the LBM are discrete
integral properties of the distribution function $f_{\alpha}$, i.e. $\sum_{\alpha=0}^8e_{\alpha x}^m e_{\alpha y}^nf_{\alpha}$,
where $m$ and $n$ are integers, we begin with the following nine non-orthogonal independent basis vectors obtained by combining
monomials $e_{\alpha x}^m e_{\alpha y}^n$ in an ascending order. That is,
\begin{eqnarray}
\ket{\rho}\equiv \ket{|\overrightarrow{e}_{\alpha}|^{0}} &=&\left(1,1,1,1,1,1,1,1,1\right)^T, \nonumber\\
\ket{e_{\alpha x}} &=&\left(0,1,0,-1,0,1,-1,-1,1\right)^T, \nonumber\\
\ket{e_{\alpha y}} &=&\left(0,0,1,0,-1,1,1,-1,-1\right)^T,\nonumber\\
\ket{e_{\alpha x}^2+e_{\alpha y}^2} &=&\left(0,1,1,1,1,2,2,2,2\right)^T,\nonumber\\
\ket{e_{\alpha x}^2-e_{\alpha y}^2} &=&\left(0,1,-1,1,-1,0,0,0,0\right)^T,\\
\ket{e_{\alpha x}e_{\alpha y}} &=&\left(0,0,0,0,0,1,-1,1,-1\right)^T,\nonumber\\
\ket{e_{\alpha x}^2e_{\alpha y}} &=&\left(0,0,0,0,0,1,1,-1,-1\right)^T,\nonumber\\
\ket{e_{\alpha x}e_{\alpha y}^2} &=&\left(0,0,0,0,0,1,-1,-1,1\right)^T,\nonumber\\
\ket{e_{\alpha x}^2e_{\alpha y}^2} &=&\left(0,0,0,0,0,1,1,1,1\right)^T,\nonumber
\end{eqnarray}
where the superscript `$T$' represents the transpose operator.

For an efficient implementation, the above non-orthogonal basis set is transformed into an equivalent orthogonal set through
the Gram-Schmidt procedure in the increasing order of the monomials of the products of the Cartesian components of the particle
velocities~\cite{premnath09d}:
\begin{eqnarray}
\ket{K_{0}}&=&\ket{\rho},\nonumber\\
\ket{K_{1}}&=&\ket{e_{\alpha x}},\nonumber\\
\ket{K_{2}}&=&\ket{e_{\alpha y}},\nonumber\\
\ket{K_{3}}&=&3\ket{e_{\alpha x}^2+e_{\alpha y}^2}-4\ket{\rho},\nonumber\\
\ket{K_{4}}&=&\ket{e_{\alpha x}^2-e_{\alpha y}^2},\\
\ket{K_{5}}&=&\ket{e_{\alpha x}e_{\alpha y}},\nonumber\\
\ket{K_{6}}&=&-3\ket{e_{\alpha x}^2e_{\alpha y}}+2\ket{e_{\alpha y}},\nonumber\\
\ket{K_{7}}&=&-3\ket{e_{\alpha x}e_{\alpha y}^2}+2\ket{e_{\alpha x}},\nonumber\\
\ket{K_{8}}&=&9\ket{e_{\alpha x}^2e_{\alpha y}^2}-6\ket{e_{\alpha x}^2+e_{\alpha y}^2}+4\ket{\rho}.\nonumber
\end{eqnarray}
This can be written explicitly in term of a matrix given by
\begin{equation}
\mathcal{K}= \left[
\begin{array}{rrrrrrrrr}
1 & 0  &  0 & -4 & 0  & 0  & 0 & 0 &  4\\
1 & 1  &  0 & -1 & 1  & 0  & 0 & 2 & -2\\
1 & 0  &  1 & -1 & -1 & 0  & 2 & 0 & -2\\
1 & -1 &  0 & -1 & 1  & 0  & 0 &-2 & -2\\
1 & 0  & -1 & -1 & -1 & 0  &-2 & 0 & -2\\
1 & 1  &  1 &  2 & 0  & 1  &-1 &-1 &  1\\
1 & -1 &  1 &  2 & 0  & -1 &-1 & 1 &  1\\
1 & -1 & -1 &  2 & 0  & 1  & 1 & 1 &  1\\
1 & 1  & -1 &  2 & 0  & -1 & 1 &-1 &  1\\
\end{array} \right],
\label{eq:collisionmatrix2}
\end{equation}
where we have used
\begin{equation}
\mathcal{K}=\left[\ket{K_{0}},\ket{K_{1}},\ket{K_{2}},\ket{K_{3}},\ket{K_{4}},\ket{K_{5}},\ket{K_{6}},\ket{K_{7}},\ket{K_{8}}\right].
\label{eq:collisionmatrix1}
\end{equation}

\section{\label{sec:continuousBoltzmannequation}Continuous Boltzmann Equation}
We consider the two-dimensional (2D) continuous Boltzmann equation, for which we aim to develop a Galilean invariant discrete model using the above basis vectors. It represents the evolution of the continuous density distribution function
$f=f(x,y,\xi_x,\xi_y)$ in continuous phase space $(x,y,\xi_x,\xi_y)$ subjected to a local force field $\overrightarrow{F}=(F_x,F_y)$,
whose origin could be internal or external to the system. By definition, the averaged effects of $f$, weighted by various powers of
the continuous particle velocity $(\xi_x,\xi_y)$, i.e. its moments, are considered to characterize the various physical processes
inherent in the motion of athermal fluids. In particular, the evolution of the slow hydrodynamical processes are described by the
local macroscopic fluid density $\rho$ and fluid velocity $\overrightarrow{u}=(u_x,u_y)$. The continuous Boltzmann equation may be
written as~\cite{chapman64,harris99}
\begin{equation}
\frac{\partial f}{\partial t}+\overrightarrow{\xi}\cdot\overrightarrow{\nabla}_{\overrightarrow{x}}f+
\frac{\overrightarrow{F}}{\rho}\cdot\overrightarrow{\nabla}_{\overrightarrow{\xi}}f=\Omega(f,f),\label{eq:continuousBoltzmannequation}
\end{equation}
where
\begin{eqnarray}
\rho &=& \int_{-\infty}^{\infty}\int_{-\infty}^{\infty}fd\xi_x d\xi_y,\\
\rho \overrightarrow{u} &=& \int_{-\infty}^{\infty}\int_{-\infty}^{\infty}f\overrightarrow{\xi}d\xi_x d\xi_y.
\end{eqnarray}
Here, $\Omega(f,f)$ is the collision term, which represents the cumulative effect of binary collision of particles. The force fields
modify the distribution function exactly by the term $-\frac{\overrightarrow{F}}{\rho}\cdot\overrightarrow{\nabla}_{\overrightarrow{\xi}}f$,
which is obtained by moving the last term on the left hand side of Eq.~(\ref{eq:continuousBoltzmannequation}) to its right to serve as
a source term. It was shown by Grad (1949)~\cite{grad49} that that solution of Eq.~(\ref{eq:continuousBoltzmannequation}) can be
approximated by the evolution equations for a hierarchical set of moments. Here, we seek to obtain a dramatically discretized
version of this continuous Boltzmann equation by means of a hierarchy of central moments, focusing, in particular, on the forcing term,
to obtain Galilean invariant representation of the dynamics of fluid motion.

\section{\label{sec:ccentralmomentsfandfeq}Continuous Central Moments: Distribution Function and its Local Attractor}
We now consider the integral properties of the distribution function $f$ given in terms of its central moments, i.e. those shifted
by the macroscopic fluid velocity. In particular, we define \emph{continuous} central moment of $f$ of order $(m+n)$ as
\begin{equation}
\widehat{\Pi}_{x^my^n}=\int_{-\infty}^{\infty}\int_{-\infty}^{\infty}f(\xi_x-u_x)^m(\xi_y-u_y)^nd\xi_xd\xi_y.
\label{eq:centralmomentfdefinition}
\end{equation}
Here, and in the rest of this paper, the use of ``hat" over a symbol represents quantities in the space of moments.
The distribution function for an athermal fluid has a local equilibrium state in the \emph{continuous} particle velocity
space $(\xi_x,\xi_y)$, which is given by the Maxwellian as~\cite{harris99}
\begin{equation}
f^\mathcal{M}\equiv
f^\mathcal{M}(\rho,\overrightarrow{u},\xi_x,\xi_y)=\frac{\rho}{2\pi c_s^2}\exp\left[-\frac{\left(\overrightarrow{\xi}-\overrightarrow{u}\right)^2}{2c_s^2}\right],\label{eq:Maxwellian}
\end{equation}
where $c_s^2=1/3$.
Analogously, we can define the corresponding central moment of the Maxwell distribution of order $(m+n)$ as
\begin{equation}
\widehat{\Pi}^{\mathcal{M}}_{x^my^n}=\int_{-\infty}^{\infty}\int_{-\infty}^{\infty}f^\mathcal{M}(\xi_x-u_x)^m(\xi_y-u_y)^nd\xi_xd\xi_y.
\label{eq:centralmomentfeqdefinition}
\end{equation}
By virtue of the fact that $f^\mathcal{M}$ being an even function, $\widehat{\Pi}^{\mathcal{M}}_{x^my^n}\neq0$ when $m$ and $n$ are even and $\widehat{\Pi}^{\mathcal{M}}_{x^my^n}=0$ when $m$ or $n$ odd. Here and henceforth, the subscripts $x^my^n$ mean
$xxx\cdots m\mbox{-times}$ and $yyy\cdots n\mbox{-times}$. Evaluation of the central moments of the Maxwellian, to different orders of increasing powers, yields
\begin{eqnarray}
\widehat{\Pi}^{\mathcal{M}}_{0}&=&\rho, \nonumber \\
\widehat{\Pi}^{\mathcal{M}}_{x}&=&0, \nonumber \\
\widehat{\Pi}^{\mathcal{M}}_{y}&=&0, \nonumber \\
\widehat{\Pi}^{\mathcal{M}}_{xx}&=&c_s^2\rho,\nonumber \\
\widehat{\Pi}^{\mathcal{M}}_{yy}&=&c_s^2\rho,\\
\widehat{\Pi}^{\mathcal{M}}_{xy}&=&0, \nonumber \\
\widehat{\Pi}^{\mathcal{M}}_{xxy}&=&0, \nonumber \\
\widehat{\Pi}^{\mathcal{M}}_{xyy}&=&0, \nonumber \\
\widehat{\Pi}^{\mathcal{M}}_{xxyy}&=&c_s^4\rho. \nonumber
\end{eqnarray}

\section{\label{sec:ccentralmomentssources}Continuous Central Moments: Forcing}
In the presence of a force field $\overrightarrow{F}$, in view of Eq.~(\ref{eq:continuousBoltzmannequation}) and as discussed in Sec.~\ref{sec:continuousBoltzmannequation}, the distribution function will be exactly modified by the source term
\begin{equation}
\delta f^{F}=-\frac{\overrightarrow{F}}{\rho}\cdot\overrightarrow{\nabla}_{\overrightarrow{\xi}}f.
\label{eq:exactsourceterm}
\end{equation}
Now we define a corresponding \emph{continuous} central moment of order $(m+n)$ due to change in the distribution function as a result of a force field as
\begin{equation}
\widehat{\Gamma}^{F}_{x^my^n}=\int_{-\infty}^{\infty}\int_{-\infty}^{\infty}\delta f^F(\xi_x-u_x)^m(\xi_y-u_y)^nd\xi_xd\xi_y.
\label{eq:centralmomentforce}
\end{equation}
Substituting Eq.~(\ref{eq:exactsourceterm}) in Eq.~(\ref{eq:centralmomentforce}) and integrating by parts by making use of the asymptotic limit assumptions $lim_{\xi_x\rightarrow \pm \infty}(\xi_x-u_x)^m f(x,y,\xi_x,\xi_y)=0$ and $lim_{\xi_y\rightarrow \pm \infty}(\xi_y-u_y)^n f(x,y,\xi_x,\xi_y)=0$, for $m,n \geq 0$, we get
\begin{eqnarray}
\widehat{\Gamma}^{F}_{x^my^n}&=&m\frac{F_x}{\rho}\int_{-\infty}^{\infty}\int_{-\infty}^{\infty}f(\xi_x-u_x)^{m-1}(\xi_y-u_y)^nd\xi_xd\xi_y+ \nonumber\\
&& n\frac{F_y}{\rho}\int_{-\infty}^{\infty}\int_{-\infty}^{\infty}f(\xi_x-u_x)^m(\xi_y-u_y)^{n-1}d\xi_xd\xi_y.
\label{eq:centralmomentforceexactidentity1}
\end{eqnarray}
From the definition given in Eq.~(\ref{eq:centralmomentfdefinition}), Eq.~(\ref{eq:centralmomentforceexactidentity1}) reduces to
an \emph{exact} identity between continuous central moment of the source term of a given order to the components of the continuous
central moment of the distribution function of an order lower acted upon by a force field:
\begin{equation}
\widehat{\Gamma}^{F}_{x^my^n}=m\frac{F_x}{\rho}\widehat{\Pi}_{x^{m-1}y^n}+ n\frac{F_y}{\rho}\widehat{\Pi}_{x^my^{n-1}},
\label{eq:centralmomentforceexactidentity2}
\end{equation}
and for the special case of the zeroth central moment of the source as $\widehat{\Gamma}^F_{0}=0$. This is a key result
based on which the rest of the derivation follows. Thus, we can enumerate the \emph{exact} values of the central moments
of sources in an increasing order as
\begin{eqnarray}
\widehat{\Gamma}^{F}_{0}&=&0, \nonumber\\
\widehat{\Gamma}^{F}_{x}&=&\frac{F_x}{\rho}\widehat{\Pi}_{0}, \nonumber\\
\widehat{\Gamma}^{F}_{y}&=&\frac{F_y}{\rho}\widehat{\Pi}_{0},\nonumber\\
\widehat{\Gamma}^{F}_{xx}&=&2\frac{F_x}{\rho}\widehat{\Pi}_{x},\nonumber\\
\widehat{\Gamma}^{F}_{yy}&=&2\frac{F_y}{\rho}\widehat{\Pi}_{y},\label{eq:exactidentitycomponents}\\
\widehat{\Gamma}^{F}_{xy}&=&\frac{F_x}{\rho}\widehat{\Pi}_{y}+\frac{F_y}{\rho}\widehat{\Pi}_{x},\nonumber\\
\widehat{\Gamma}^{F}_{xxy}&=&2\frac{F_x}{\rho}\widehat{\Pi}_{xy}+\frac{F_y}{\rho}\widehat{\Pi}_{xx},\nonumber\\
\widehat{\Gamma}^{F}_{xyy}&=&\frac{F_x}{\rho}\widehat{\Pi}_{yy}+2\frac{F_y}{\rho}\widehat{\Pi}_{xy},\nonumber\\
\widehat{\Gamma}^{F}_{xxyy}&=&2\frac{F_x}{\rho}\widehat{\Pi}_{xyy}+2\frac{F_y}{\rho}\widehat{\Pi}_{xxy}.\nonumber
\end{eqnarray}
Note that if we set $\widehat{\Pi}_{x^my^n}=\widehat{\Pi}_{x^my^n}^{\mathcal{M}}$ in Eq.~(\ref{eq:exactidentitycomponents}), i.e. ignore non-equilibrium effects, we arrive at the the derivation given in~\cite{premnath09d} as a special case.

\section{\label{sec:cascadedLBEforcing}Cascaded Central Moment Lattice-Boltzmann Method with Forcing Terms}
Defining a \emph{discrete} distribution function supported by the discrete particle velocity set $\overrightarrow{e}_{\alpha}$ as
$\mathbf{f}=\ket{f_{\alpha}}=(f_0,f_1,f_2,\ldots,f_8)^T$, and a cascaded collision operator as $\bm{\Omega}^{c}=\ket{\Omega_{\alpha}^{c}}=(\Omega_{0}^{c},\Omega_{1}^{c},\Omega_{2}^{c},\ldots,\Omega_{8}^{c})^T$ as well as a source term as $\mathbf{S}=\ket{S_{\alpha}}=(S_0,S_1,S_2,\ldots,S_8)^T$ based on central moments,
we obtain the lattice Boltzmann equation (LBE) as a discrete version of Eq.~(\ref{eq:continuousBoltzmannequation})
by temporally integrating along particle characteristics as follows~\cite{premnath09d}:
\begin{equation}
f_{\alpha}(\overrightarrow{x}+\overrightarrow{e}_{\alpha},t+1)=f_{\alpha}(\overrightarrow{x},t)+\Omega_{{\alpha}(\overrightarrow{x},t)}^{c}+
\int_{t}^{t+1}S_{{\alpha}(\overrightarrow{x}+\overrightarrow{e}_{\alpha}\theta,t+\theta)}d\theta.
\label{eq:cascadedLBE1}
\end{equation}
Here, the collision operator is written as
\begin{equation}
\Omega_{\alpha}^{c}\equiv \Omega_{\alpha}^{c}(\mathbf{f},\mathbf{\widehat{g}})=(\mathcal{K}\cdot \mathbf{\widehat{g}})_{\alpha},
\label{eq:cascadecollision1}
\end{equation}
where $\mathbf{\widehat{g}}=\ket{\widehat{g}_{\alpha}}=(\widehat{g}_0,\widehat{g}_1,\widehat{g}_2,\ldots,\widehat{g}_8)^T$.
The hydrodynamic fields are obtained from the distribution function as
\begin{eqnarray}
\rho&=&\sum_{\alpha=0}^{8}f_{\alpha}=\braket{f_{\alpha}|\rho},\\
\rho u_i&=&\sum_{\alpha=0}^{8}f_{\alpha} e_{\alpha i}=\braket{f_{\alpha}|e_{\alpha i}}, i \in {x,y}.
\end{eqnarray}
For improved accuracy in recovering Navier-Stokes solution, using a semi-implicit representation for the source term, i.e. the last term in the above equation (Eq.~(\ref{eq:cascadedLBE1})) as $\int_{t}^{t+1}S_{{\alpha}(\overrightarrow{x}+\overrightarrow{e}_{\alpha}\theta,t+\theta)}d\theta=\frac{1}{2}\left[S_{{\alpha}(\overrightarrow{x},t)}+S_{{\alpha}(\overrightarrow{x}+\overrightarrow{e}_{\alpha},t+1)}\right]$, so that Eq.~(\ref{eq:cascadedLBE1}) is made effectively explicit by using the transformation
$\overline{f}_{\alpha}=f_{\alpha}-\frac{1}{2}S_{\alpha}$ to reduce it to~\cite{premnath09d}
\begin{equation}
\overline{f}_{\alpha}(\overrightarrow{x}+\overrightarrow{e}_{\alpha},t+1)=\overline{f}_{\alpha}(\overrightarrow{x},t)+\Omega_{{\alpha}(\overrightarrow{x},t)}^{c}+
S_{{\alpha}(\overrightarrow{x},t)}.
\label{eq:cascadedLBE3}
\end{equation}
It may be noted that, as in~\cite{geier06}, we first represent collision as a cascaded process in which the effect of collision
on lower order central moments successively influence those at higher orders in a cascaded manner. That is, in general,
$\widehat{g}_{\alpha}\equiv\widehat{g}_{\alpha}(\mathbf{f},\widehat{g}_{\beta}), \beta=0,1,2,\ldots,\alpha-1$.
Furthermore, the form of the source term is derived to rigorously enforce Galilean invariance.
The explicit expressions for $S_{\alpha}$ and $\mathbf{\widehat{g}}$ will be determined later
in Secs.~\ref{sec:rawmoments} and ~\ref{sec:cascadedcollisionforcing}, respectively. Since the main focus of this work is on improving the collision (including forcing) step with features independent of inertial frames, we have only considered the standard discretization for the streaming operator. However, as discussed in the Introduction, other types of discretization schemes could be considered to improve the order of accuracy.

\section{\label{sec:discretecentralmoments}Various Discrete Central Moments and Galilean Invariance Matching Principle}
For determining the structure of the cascaded collision operator $\mathbf{\widehat{g}}$ and the source terms $S_{\alpha}$, we first
need to define the following \emph{discrete} central moments of the distribution function, Maxwellian, and source term, respectively:
\begin{eqnarray}
\widehat{\kappa}_{x^m y^n}&=&\sum_{\alpha}f_{\alpha}(e_{\alpha x}-u_x)^m(e_{\alpha y}-u_y)^n=\braket{(e_{\alpha x}-u_x)^m(e_{\alpha y}-u_y)^n|f_{\alpha}},\label{eq:centralmomentdistributionfunction1}\\
\widehat{\kappa}_{x^m y^n}^{\mathcal{M}}&=&\sum_{\alpha}f_{\alpha}^{\mathcal{M}}(e_{\alpha x}-u_x)^m(e_{\alpha y}-u_y)^n=\braket{(e_{\alpha x}-u_x)^m(e_{\alpha y}-u_y)^n|f_{\alpha}^{\mathcal{M}}},\label{eq:centralmomentMaxwelldistribution1}\\
\widehat{\sigma}_{x^m y^n}&=&\sum_{\alpha}S_{\alpha}(e_{\alpha x}-u_x)^m(e_{\alpha y}-u_y)^n=\braket{(e_{\alpha x}-u_x)^m(e_{\alpha y}-u_y)^n|S_{\alpha}}\label{eq:centralmomentforcingterm1}.
\end{eqnarray}
where the exact expression for the discrete $f_{\alpha}^{\mathcal{M}}$ is not yet known, but can be determined as a result of the
derivation discussed later.
To maintain physical consistency at the discrete level, we now equate the \emph{discrete} central moments of the distribution function,
the Maxwellian and the source terms, defined above, with their corresponding \emph{continuous} central moments, whose forms are
known exactly. That is, according to this matching principle
\begin{eqnarray}
\widehat{\kappa}_{x^m y^n}&=&\widehat{\Pi}_{x^m y^n},\\
\widehat{\kappa}_{x^m y^n}^{\mathcal{M}}&=&\widehat{\Pi}^{\mathcal{M}}_{x^m y^n},\\
\widehat{\sigma}_{x^m y^n}&=&\widehat{\Gamma}^{F}_{x^m y^n}.
\end{eqnarray}
In particular, the discrete central moments of various orders for both the Maxwellian and the source terms, respectively, become
\begin{eqnarray}
\widehat{\kappa}^{\mathcal{M}}_{0}&=&\rho, \nonumber\\
\widehat{\kappa}^{\mathcal{M}}_{x}&=&0, \nonumber\\
\widehat{\kappa}^{\mathcal{M}}_{y}&=&0, \nonumber\\
\widehat{\kappa}^{\mathcal{M}}_{xx}&=&c_s^2\rho,\nonumber\\
\widehat{\kappa}^{\mathcal{M}}_{yy}&=&c_s^2\rho,\label{eq:centralmomentMaxwelldistribution}\\
\widehat{\kappa}^{\mathcal{M}}_{xy}&=&0, \nonumber\\
\widehat{\kappa}^{\mathcal{M}}_{xxy}&=&0, \nonumber\\
\widehat{\kappa}^{\mathcal{M}}_{xyy}&=&0, \nonumber\\
\widehat{\kappa}^{\mathcal{M}}_{xxyy}&=&c_s^4\rho, \nonumber
\end{eqnarray}
and
\begin{eqnarray}
\widehat{\sigma}_{0}&=&0, \nonumber\\
\widehat{\sigma}_{x}&=&\frac{F_x}{\rho}\widehat{\kappa}_{0}, \nonumber\\
\widehat{\sigma}_{y}&=&\frac{F_y}{\rho}\widehat{\kappa}_{0}, \nonumber\\
\widehat{\sigma}_{xx}&=&2\frac{F_x}{\rho}\widehat{\kappa}_{x},\nonumber\\
\widehat{\sigma}_{yy}&=&2\frac{F_y}{\rho}\widehat{\kappa}_{y},\label{eq:centralmomentforcing}\\
\widehat{\sigma}_{xy}&=&\frac{F_x}{\rho}\widehat{\kappa}_{y}+\frac{F_y}{\rho}\widehat{\kappa}_{x},\nonumber\\
\widehat{\sigma}_{xxy}&=&2\frac{F_x}{\rho}\widehat{\kappa}_{xy}+\frac{F_y}{\rho}\widehat{\kappa}_{xx},\nonumber\\
\widehat{\sigma}_{xyy}&=&\frac{F_x}{\rho}\widehat{\kappa}_{yy}+2\frac{F_y}{\rho}\widehat{\kappa}_{xy},\nonumber\\
\widehat{\sigma}_{xxyy}&=&2\frac{F_x}{\rho}\widehat{\kappa}_{xyy}+2\frac{F_y}{\rho}\widehat{\kappa}_{xxy}.\nonumber
\end{eqnarray}
We also define a \emph{discrete} central moment in terms of the transformed distribution function to facilitate subsequent
developments as
\begin{equation}
\widehat{\overline{\kappa}}_{x^m y^n}=\sum_{\alpha}\overline{f}_{\alpha}(e_{\alpha x}-u_x)^m(e_{\alpha y}-u_y)^n=\braket{(e_{\alpha x}-u_x)^m(e_{\alpha y}-u_y)^n|\overline{f}_{\alpha}}.
\end{equation}
Owing to the transformation discussed in Sec.~\ref{sec:cascadedLBEforcing}, it follows that
\begin{equation}
\widehat{\overline{\kappa}}_{x^m y^n}=\widehat{\kappa}_{x^m y^n}-\frac{1}{2}\widehat{\sigma}_{x^m y^n}.
\label{eq:transformedcentralmoments}
\end{equation}

\section{\label{sec:discretemomentsrelation}Relation Between Various Discrete Central Moments}
Equation~(\ref{eq:centralmomentforcing}) is given in terms of the discrete moments of the original distribution function $f_{\alpha}$. However, the cascaded central moment LBM with forcing term provides evolution in terms of transformed distribution function $\overline{f}_{\alpha}$ (Eq.~(\ref{eq:cascadedLBE3})). Thus, it is important to write all the expressions in terms of the central moments of $\overline{f}_{\alpha}$, or, equivalently, $\widehat{\overline{\kappa}}_{x^m y^n}$. Thus, by recursive application of Eq.~(\ref{eq:transformedcentralmoments}) using Eq.~(\ref{eq:centralmomentforcing}) to successively higher orders, we get the following exact relations up to the third-order central moments as
\begin{eqnarray}
\widehat{\kappa}_{0}&=&\widehat{\overline{\kappa}}_{0},\nonumber\\
\widehat{\kappa}_{x}&=&\widehat{\overline{\kappa}}_{x}+\frac{1}{2}\frac{F_x}{\rho}\widehat{\overline{\kappa}}_{0},\nonumber\\
\widehat{\kappa}_{y}&=&\widehat{\overline{\kappa}}_{y}+\frac{1}{2}\frac{F_y}{\rho}\widehat{\overline{\kappa}}_{0},\nonumber\\
\widehat{\kappa}_{xx}&=&\widehat{\overline{\kappa}}_{xx}+\frac{F_x}{\rho}\widehat{\overline{\kappa}}_{x}+\frac{1}{2}\frac{F_x^2}{\rho^2}\widehat{\overline{\kappa}}_{0},\nonumber\\
\widehat{\kappa}_{yy}&=&\widehat{\overline{\kappa}}_{yy}+\frac{F_x}{\rho}\widehat{\overline{\kappa}}_{y}+\frac{1}{2}\frac{F_y^2}{\rho^2}\widehat{\overline{\kappa}}_{0},\nonumber\\
\widehat{\kappa}_{xy}&=&\widehat{\overline{\kappa}}_{xy}+\frac{1}{2}\frac{F_x}{\rho}\widehat{\overline{\kappa}}_{y}+\frac{1}{2}\frac{F_y}{\rho}\widehat{\overline{\kappa}}_{x}+\frac{1}{2}\frac{F_xF_y}{\rho^2}\widehat{\overline{\kappa}}_{0},\nonumber\\
\widehat{\kappa}_{xxy}&=&\widehat{\overline{\kappa}}_{xxy}+\frac{F_x}{\rho}\widehat{\overline{\kappa}}_{xy}+
\frac{1}{2}\frac{F_y}{\rho}\widehat{\overline{\kappa}}_{xx}+\frac{F_xF_y}{\rho^2}\widehat{\overline{\kappa}}_{x}+\frac{1}{2}\frac{F_x^2}{\rho^2}\widehat{\overline{\kappa}}_{y}
+\frac{3}{4}\frac{F_x^2F_y}{\rho^3}\widehat{\overline{\kappa}}_{0},\nonumber\\
\widehat{\kappa}_{xyy}&=&\widehat{\overline{\kappa}}_{xyy}+\frac{F_y}{\rho}\widehat{\overline{\kappa}}_{xy}+
\frac{1}{2}\frac{F_x}{\rho}\widehat{\overline{\kappa}}_{yy}+\frac{F_xF_y}{\rho^2}\widehat{\overline{\kappa}}_{y}+\frac{1}{2}\frac{F_y^2}{\rho^2}\widehat{\overline{\kappa}}_{x}
+\frac{3}{4}\frac{F_xF_y^2}{\rho^3}\widehat{\overline{\kappa}}_{0}.\nonumber
\end{eqnarray}
That is, the central moment of the distribution function of a given order can be written as a function of the central moment of the
transformed distribution function of the same order and successively lower orders as well. This can be further simplified by
considering the three of the lowest order central moments, i.e., conservative moments, which by definition are $\widehat{\kappa}_{0}=\rho$,
$\widehat{\kappa}_{x}=\widehat{\kappa}_{y}=0$. This, in turn, leads to $\widehat{\overline{\kappa}}_{0}=\rho$,
$\widehat{\overline{\kappa}}_{x}=-1/2F_x$, $\widehat{\overline{\kappa}}_{y}=-1/2F_y$. As a result, we have the following relations for the non-conserved central moments up to third-order:
\begin{eqnarray}
\widehat{\kappa}_{xx}&=&\widehat{\overline{\kappa}}_{xx}, \nonumber\\
\widehat{\kappa}_{yy}&=&\widehat{\overline{\kappa}}_{yy}, \nonumber\\
\widehat{\kappa}_{xy}&=&\widehat{\overline{\kappa}}_{xy}, \nonumber\\
\widehat{\kappa}_{xxy}&=&\widehat{\overline{\kappa}}_{xxy}+\frac{F_x}{\rho}\widehat{\overline{\kappa}}_{xy}+
\frac{1}{2}\frac{F_y}{\rho}\widehat{\overline{\kappa}}_{xx}, \nonumber\\
\widehat{\kappa}_{xyy}&=&\widehat{\overline{\kappa}}_{xyy}+\frac{F_y}{\rho}\widehat{\overline{\kappa}}_{xy}+
\frac{1}{2}\frac{F_x}{\rho}\widehat{\overline{\kappa}}_{yy}. \nonumber
\end{eqnarray}
Thus, we can finally write the central moments of the source term in Eq.~(\ref{eq:centralmomentforcing})
in terms of the central moments of the transformed distribution function as
\begin{eqnarray}
\widehat{\sigma}_{0}&=&0, \nonumber\\
\widehat{\sigma}_{x}&=&F_x, \nonumber\\
\widehat{\sigma}_{y}&=&F_y, \nonumber\\
\widehat{\sigma}_{xx}&=&0, \nonumber\\
\widehat{\sigma}_{yy}&=&0, \label{eq:centralmomentforcingtransformed}\\
\widehat{\sigma}_{xy}&=&0, \nonumber\\
\widehat{\sigma}_{xxy}&=&2\frac{F_x}{\rho}\widehat{\overline{\kappa}}_{xy}+\frac{F_y}{\rho}\widehat{\overline{\kappa}}_{xx}, \nonumber\\
\widehat{\sigma}_{xyy}&=&\frac{F_x}{\rho}\widehat{\overline{\kappa}}_{yy}+2\frac{F_y}{\rho}\widehat{\overline{\kappa}}_{xy}, \nonumber\\
\widehat{\sigma}_{xxyy}&=&2\frac{F_x}{\rho}\widehat{\overline{\kappa}}_{xyy}+
2\frac{F_y}{\rho}\widehat{\overline{\kappa}}_{xxy}+\frac{F_y^2}{\rho^2}\widehat{\overline{\kappa}}_{xx}+
\frac{F_x^2}{\rho^2}\widehat{\overline{\kappa}}_{yy}+4\frac{F_xF_y}{\rho^2}\widehat{\overline{\kappa}}_{xy}. \nonumber
\end{eqnarray}
Thus, higher order non-equilibrium effects in $\widehat{\overline{\kappa}}_{x^m y^n}$ and non-linear effect in $F_x^pF_y^q$
are evident for the central moments of the source terms that are third- and higher orders. Let us now explicitly write the central moments of the transformed discrete Maxwellian by means of Eq.~(\ref{eq:transformedcentralmoments}) using Eqs.~(\ref{eq:centralmomentMaxwelldistribution}) and (\ref{eq:centralmomentforcingtransformed}) to yield
\begin{eqnarray}
\widehat{\overline{\kappa}}^{\mathcal{M}}_{0}&=&\rho, \nonumber\\
\widehat{\overline{\kappa}}^{\mathcal{M}}_{x}&=&-\frac{1}{2}F_x, \nonumber\\
\widehat{\overline{\kappa}}^{\mathcal{M}}_{y}&=&-\frac{1}{2}F_y, \nonumber\\
\widehat{\overline{\kappa}}^{\mathcal{M}}_{xx}&=&c_s^2\rho, \nonumber\\
\widehat{\overline{\kappa}}^{\mathcal{M}}_{yy}&=&c_s^2\rho, \label{eq:centralmomenttransformed}\\
\widehat{\overline{\kappa}}^{\mathcal{M}}_{xy}&=&0, \nonumber\\
\widehat{\overline{\kappa}}^{\mathcal{M}}_{xxy}&=&-\frac{F_x}{\rho}\widehat{\overline{\kappa}}_{xy}-\frac{F_y}{2\rho}\widehat{\overline{\kappa}}_{xx}, \nonumber\\
\widehat{\overline{\kappa}}^{\mathcal{M}}_{xyy}&=&-\frac{F_x}{2\rho}\widehat{\overline{\kappa}}_{yy}-\frac{F_y}{\rho}\widehat{\overline{\kappa}}_{xy}, \nonumber\\
\widehat{\overline{\kappa}}^{\mathcal{M}}_{xxyy}&=&c_s^4\rho-\frac{F_x}{\rho}\widehat{\overline{\kappa}}_{xyy}-
\frac{F_y}{\rho}\widehat{\overline{\kappa}}_{xxy}-\frac{F_y^2}{2\rho^2}\widehat{\overline{\kappa}}_{xx}-
\frac{F_x^2}{2\rho^2}\widehat{\overline{\kappa}}_{yy}-2\frac{F_xF_y}{\rho^2}\widehat{\overline{\kappa}}_{xy}.\nonumber
\end{eqnarray}
The main idea in the determination of the collision operator for the cascaded version of the central moment method is to relax the
central moments of the transformed distribution function to its corresponding local attractor, successively at various orders as given in Eq.~(\ref{eq:centralmomenttransformed}) (see Sec.~\ref{sec:cascadedcollisionforcing}). Before proceeding further to do this,
we first need certain quantities in the rest or lattice frame of reference, i.e. the raw moments, in which the computations are
actually performed. These are obtained in the next section.

\section{\label{sec:rawmoments}Various Discrete Raw Moments and Source Terms in Particle Velocity Space}
The raw moments, i.e. those in the rest frame of reference, can be related to the central moments by means of
the binomial theorem~\cite{kogan69,muller93}. For any state variable $\varphi$ supported by the discrete particle
velocity set, the transformation relation between the two reference frames is thus given by~\cite{premnath09d}
\begin{eqnarray}
\braket{(e_{\alpha x}-u_x)^m(e_{\alpha y}-u_y)^n|\varphi}&=&\braket{e_{\alpha x}^m e_{\alpha y}^n|\varphi}+
\braket{e_{\alpha x}^m \left[\sum_{j=1}^{n} C^n_je_{\alpha y}^{n-j}(-1)^{j}u_y^{j}\right]|\varphi}\nonumber +\\
&& \braket{e_{\alpha y}^n \left[\sum_{i=1}^{m} C^m_ie_{\alpha x}^{m-i}(-1)^{i}u_x^{i}\right]|\varphi}\nonumber +\\
&& \braket{ \left[\sum_{i=1}^{m} C^m_ie_{\alpha x}^{m-i}(-1)^{i}u_x^{i}\right]\left[\sum_{j=1}^{n} C^n_je_{\alpha y}^{n-j}(-1)^{j}u_y^{j}\right]|\varphi}
\label{eq:binomialtheorem}
\end{eqnarray}
where
$C^{p}_{q}=p!/(q!(p-q)!)$.
We now define the following notations for depicting various \emph{discrete raw} moments, based on which an operational LBE will
be devised later:
\begin{eqnarray}
\widehat{\kappa}_{x^m y^n}^{'}&=&\sum_{\alpha}f_{\alpha}e_{\alpha x}^m e_{\alpha y}^n=\braket{e_{\alpha x}^m e_{\alpha y}^n|f_{\alpha}},\label{eq:rawmomentdistributionfunction1}\\
\widehat{\overline{\kappa}}_{x^m y^n}^{'}&=&\sum_{\alpha}\overline{f}_{\alpha}e_{\alpha x}^m e_{\alpha y}^n=\braket{e_{\alpha x}^m e_{\alpha y}^n|\overline{f}_{\alpha}},\label{eq:rawmomenttransformeddistribution1}\\
\widehat{\sigma}_{x^m y^n}^{'}&=&\sum_{\alpha}S_{\alpha}e_{\alpha x}^m e_{\alpha y}^n=\braket{e_{\alpha x}^m e_{\alpha y}^n|S_{\alpha}}\label{eq:rawmomentforcingterm1}.
\end{eqnarray}
Note that the superscript ``prime'' ($'$) is used to distinguish the raw moments from the central moments that are designated without
the primes. Here, analogous to Eq.~(\ref{eq:transformedcentralmoments}), the relation $\widehat{\overline{\kappa}}_{x^m y^n}^{'}=\widehat{\kappa}_{x^m y^n}^{'}-\frac{1}{2}\widehat{\sigma}_{x^m y^n}^{'}$ holds. Let us
first find expressions for $\widehat{\overline{\kappa}}_{x^m y^n}^{'}=\braket{\overline{f}_{\alpha}|e_{\alpha x}^m e_{\alpha y}^n}$
to proceed further. As in~\cite{premnath09d}, for convenience, we define the following operator acting on the transformed distribution
function $\overline{f}_{\alpha}$ in this regard:
\begin{equation}
a(\overline{f}_{\alpha_1}+\overline{f}_{\alpha_3}+\overline{f}_{\alpha_3}+\cdots)+
b(\overline{f}_{\beta_1}+\overline{f}_{\beta_2}+\overline{f}_{\beta_3}+\cdots)+\cdots=\left(a\sum_{\alpha}^A+b\sum_{\alpha}^B\cdots\right)\otimes \overline{f}_{\alpha},\label{eq:summationoperator}
\end{equation}
where $A=\left\{\alpha_1,\alpha_2,\alpha_3,\cdots\right\}$,~$B=\left\{\beta_1,\beta_2,\beta_3,\cdots\right\}$,$\cdots$.
For conserved basis vectors, we write them in terms of the hydrodynamic variables and force fields as
\begin{eqnarray}
\widehat{\overline{\kappa}}_{0}^{'}&=&\braket{\overline{f}_{\alpha}|\rho}=\sum_{\alpha = 0}^{8}\overline{f}_{\alpha}=\rho,\label{eq:rawmomenttransformed0}\\
\widehat{\overline{\kappa}}_{x}^{'}&=&\braket{\overline{f}_{\alpha}|e_{\alpha x}}=\sum_{\alpha = 0}^{8}\overline{f}_{\alpha} e_{\alpha x}=\rho u_x-\frac{1}{2}F_x,\label{eq:rawmomenttransformed1}\\
\widehat{\overline{\kappa}}_{y}^{'}&=&\braket{\overline{f}_{\alpha}|e_{\alpha y}}=\sum_{\alpha = 0}^{8}\overline{f}_{\alpha} e_{\alpha y}=\rho u_y-\frac{1}{2}F_y,\label{eq:rawmomenttransformed2}
\end{eqnarray}
and, for the non-conserved basis vectors, using Eq.~(\ref{eq:summationoperator}) in terms of subsets of particle velocity directions as
\begin{eqnarray}
\widehat{\overline{\kappa}}_{xx}^{'}&=&\braket{\overline{f}_{\alpha}|e_{\alpha x}^2}=\sum_{\alpha = 0}^{8}\overline{f}_{\alpha} e_{\alpha x}^2=\left(\sum_{\alpha}^{A_3}\right)\otimes \overline{f}_{\alpha},\label{eq:rawmomenttransformed3}\\
\widehat{\overline{\kappa}}_{yy}^{'}&=&\braket{\overline{f}_{\alpha}|e_{\alpha y}^2}=\sum_{\alpha = 0}^{8}\overline{f}_{\alpha} e_{\alpha y}^2=\left(\sum_{\alpha}^{A_4}\right)\otimes \overline{f}_{\alpha},\label{eq:rawmomenttransformed4}\\
\widehat{\overline{\kappa}}_{xy}^{'}&=&\braket{\overline{f}_{\alpha}|e_{\alpha x}e_{\alpha y}}=\sum_{\alpha = 0}^{8}\overline{f}_{\alpha} e_{\alpha x} e_{\alpha y}=\left(\sum_{\alpha}^{A_5}-\sum_{\alpha}^{B_5}\right)\otimes \overline{f}_{\alpha},\label{eq:rawmomenttransformed5}\\
\widehat{\overline{\kappa}}_{xxy}^{'}&=&\braket{\overline{f}_{\alpha}|e_{\alpha x}^2e_{\alpha y}}=\sum_{\alpha = 0}^{8}\overline{f}_{\alpha} e_{\alpha x}^2 e_{\alpha y}=\left(\sum_{\alpha}^{A_6}-\sum_{\alpha}^{B_6}\right)\otimes \overline{f}_{\alpha},\label{eq:rawmomenttransformed6}\\
\widehat{\overline{\kappa}}_{xyy}^{'}&=&\braket{\overline{f}_{\alpha}|e_{\alpha x}e_{\alpha y}^2}=\sum_{\alpha = 0}^{8}\overline{f}_{\alpha} e_{\alpha x} e_{\alpha y}^2=\left(\sum_{\alpha}^{A_7}-\sum_{\alpha}^{B_7}\right)\otimes \overline{f}_{\alpha},\label{eq:rawmomenttransformed7}\\
\widehat{\overline{\kappa}}_{xxyy}^{'}&=&\braket{\overline{f}_{\alpha}|e_{\alpha x}^2e_{\alpha y}^2}=\sum_{\alpha = 0}^{8}\overline{f}_{\alpha} e_{\alpha x}^2 e_{\alpha y}^2=\left(\sum_{\alpha}^{A_8}\right)\otimes \overline{f}_{\alpha},\label{eq:rawmomenttransformed8}
\end{eqnarray}
where
\begin{eqnarray*}
A_3&=&\left\{1,3,5,6,7,8\right\},\\
A_4&=&\left\{2,4,5,6,7,8\right\},\\
A_5&=&\left\{5,7\right\},B_5=\left\{6,8\right\},\\
A_6&=&\left\{5,6\right\},B_6=\left\{7,8\right\},\\
A_7&=&\left\{5,8\right\},B_7=\left\{6,7\right\},\\
A_8&=&\left\{5,6,7,8\right\}.
\end{eqnarray*}

Now, we transform the central moments of the source terms (Eq.~(\ref{eq:centralmomentforcing})) to
the corresponding raw moments by considering Eq.~(\ref{eq:centralmomentforcingterm1}) and using the frame transformation
relation (Eq.~(\ref{eq:binomialtheorem})). This yields
\begin{eqnarray}
&&\widehat{\sigma}_{0}^{'}=\braket{S_{\alpha}|\rho}=0,\label{eq:rawmomentsourceterm0}\\
&&\widehat{\sigma}_{x}^{'}=\braket{S_{\alpha}|e_{\alpha x}}=F_x,\label{eq:rawmomentsourceterm1}\\
&&\widehat{\sigma}_{y}^{'}=\braket{S_{\alpha}|e_{\alpha y}}=F_y,\label{eq:rawmomentsourceterm2}\\
&&\widehat{\sigma}_{xx}^{'}=\braket{S_{\alpha}|e_{\alpha x}^2}=2F_xu_x,\label{eq:rawmomentsourceterm3}\\
&&\widehat{\sigma}_{yy}^{'}=\braket{S_{\alpha}|e_{\alpha y}^2}=2F_xu_y,\label{eq:rawmomentsourceterm4}\\
&&\widehat{\sigma}_{xy}^{'}=\braket{S_{\alpha}|e_{\alpha x}e_{\alpha y}}=F_xu_y+F_yu_x,\label{eq:rawmomentsourceterm5}\\
&&\widehat{\sigma}_{xxy}^{'}=\braket{S_{\alpha}|e_{\alpha x}^2e_{\alpha y}}=2F_x\left(\frac{\widehat{\overline{\kappa}}_{xy}}{\rho}+u_xu_y\right)+F_y\left(\frac{\widehat{\overline{\kappa}}_{xx}}{\rho}+u_x^2\right),\label{eq:rawmomentsourceterm6}\\
&&\widehat{\sigma}_{xyy}^{'}=\braket{S_{\alpha}|e_{\alpha x}e_{\alpha y}^2}=F_x\left(\frac{\widehat{\overline{\kappa}}_{yy}}{\rho}+u_y^2\right)+2F_y\left(\frac{\widehat{\overline{\kappa}}_{xy}}{\rho}+u_xu_y\right),\label{eq:rawmomentsourceterm7}\\
&&\widehat{\sigma}_{xxyy}^{'}=\braket{S_{\alpha}|e_{\alpha x}^2e_{\alpha y}^2}=2F_xu_x\left(\frac{\widehat{\overline{\kappa}}_{yy}}{\rho}+u_y^2\right)+2F_yu_y\left(\frac{\widehat{\overline{\kappa}}_{xx}}{\rho}+u_x^2\right)+\nonumber\\
&&\frac{2F_x}{\rho}\widehat{\overline{\kappa}}_{xyy}+
\frac{2F_y}{\rho}\widehat{\overline{\kappa}}_{xxy}+
\frac{F_x^2}{\rho^2}\widehat{\overline{\kappa}}_{yy}+\frac{F_y^2}{\rho^2}\widehat{\overline{\kappa}}_{xx}+\nonumber\\
&&4\left[\frac{F_xu_y}{\rho}+\frac{F_yu_x}{\rho}+\frac{F_xF_y}{\rho^2}\right]\widehat{\overline{\kappa}}_{xy}.\label{eq:rawmomentsourceterm8}
\end{eqnarray}
Clearly, the raw moments of source terms for third-order or higher contain non-equilibrium and non-linear contributions.
Eqs.~(\ref{eq:rawmomentsourceterm6})-(\ref{eq:rawmomentsourceterm8}) require explicit expressions for central moments of
transformed distributions such as $\widehat{\overline{\kappa}}_{xx}$, $\widehat{\overline{\kappa}}_{yy}$,
$\widehat{\overline{\kappa}}_{xy}$, $\widehat{\overline{\kappa}}_{xxy}$ and $\widehat{\overline{\kappa}}_{xyy}$,
in terms of raw moments to facilitate computation. They can be readily obtained in terms of raw moments from their respective
definitions and by using the binomial theorem (Eq.~(\ref{eq:binomialtheorem})) and subsequent simplification as
\begin{eqnarray}
\widehat{\overline{\kappa}}_{xx}&=&\widehat{\overline{\kappa}}_{xx}^{'}+F_xu_x-\rho u_x^2, \label{eq:kxxtransform}\\
\widehat{\overline{\kappa}}_{yy}&=&\widehat{\overline{\kappa}}_{yy}^{'}+F_yu_y-\rho u_y^2, \label{eq:kyytransform}\\
\widehat{\overline{\kappa}}_{xy}&=&\widehat{\overline{\kappa}}_{xy}^{'}+\frac{1}{2}(F_xu_y+F_yu_x)-\rho u_xu_y,\label{eq:kxytransform}
\end{eqnarray}
for second-order and
\begin{eqnarray}
\widehat{\overline{\kappa}}_{xyy}&=&\widehat{\overline{\kappa}}_{xyy}^{'}-2u_y\widehat{\overline{\kappa}}_{xy}^{'}-u_x\widehat{\overline{\kappa}}_{yy}^{'}-
\frac{1}{2}F_xu_y^2-F_yu_xu_y+2\rho u_x u_y^2,\label{eq:kxyytransform}\\
\widehat{\overline{\kappa}}_{xxy}&=&\widehat{\overline{\kappa}}_{xxy}^{'}-2u_x\widehat{\overline{\kappa}}_{xy}^{'}-u_y\widehat{\overline{\kappa}}_{xx}^{'}-
\frac{1}{2}F_yu_x^2-F_xu_xu_y+2\rho u_x^2 u_y,\label{eq:kxxytransform}
\end{eqnarray}
for third-order moments. Based on the above, we now obtain the source terms projected to the orthogonal moment basis vectors, i.e.
$\braket{K_{\beta}|S_{\alpha}}$, $\beta=0,1,2,\ldots,8$, which would then provide corresponding explicit expressions
in terms of the particle velocity space. Thus, from Eqs.~(\ref{eq:collisionmatrix1}) and (\ref{eq:rawmomentsourceterm0})-(\ref{eq:rawmomentsourceterm8}),
the following projected source moments are derived:
\begin{eqnarray}
\widehat{m}^{s}_{0}=\braket{K_0|S_{\alpha}}&=&0, \label{eq:sourceterm0}\\
\widehat{m}^{s}_{1}=\braket{K_1|S_{\alpha}}&=&F_x, \label{eq:sourceterm1}\\
\widehat{m}^{s}_{2}=\braket{K_2|S_{\alpha}}&=&F_y, \label{eq:sourceterm2}\\
\widehat{m}^{s}_{3}=\braket{K_3|S_{\alpha}}&=&6(F_xu_x+F_yu_y), \label{eq:sourceterm3}\\
\widehat{m}^{s}_{4}=\braket{K_4|S_{\alpha}}&=&2(F_xu_x-F_yu_y), \label{eq:sourceterm4}\\
\widehat{m}^{s}_{5}=\braket{K_5|S_{\alpha}}&=&(F_xu_y+F_yu_x), \label{eq:sourceterm5}\\
\widehat{m}^{s}_{6}=\braket{K_6|S_{\alpha}}&=&-6F_x\left(\frac{\widehat{\overline{\kappa}}_{xy}}{\rho}+u_xu_y\right)-
3F_y\left(\frac{\widehat{\overline{\kappa}}_{xx}}{\rho}+u_x^2-\frac{2}{3}\right), \label{eq:sourceterm6}\\
\widehat{m}^{s}_{7}=\braket{K_7|S_{\alpha}}&=&-3F_x\left(\frac{\widehat{\overline{\kappa}}_{yy}}{\rho}+u_y^2-\frac{2}{3}\right)-
6F_y\left(\frac{\widehat{\overline{\kappa}}_{xy}}{\rho}+u_xu_y\right), \label{eq:sourceterm7}\\
\widehat{m}^{s}_{8}=\braket{K_8|S_{\alpha}}&=&18F_xu_x\left(\frac{\widehat{\overline{\kappa}}_{yy}}{\rho}+u_y^2-\frac{2}{3}\right)+
18F_yu_y\left(\frac{\widehat{\overline{\kappa}}_{xx}}{\rho}+u_x^2-\frac{2}{3}\right)+\nonumber\\
&&18\frac{F_x}{\rho}\widehat{\overline{\kappa}}_{xyy}+18\frac{F_y}{\rho}\widehat{\overline{\kappa}}_{xxy}+
9\frac{F_x^2}{\rho^2}\widehat{\overline{\kappa}}_{yy}+9\frac{F_y^2}{\rho^2}\widehat{\overline{\kappa}}_{xx}+\nonumber\\
&&
36\left[\frac{F_xu_y}{\rho}+\frac{F_yu_x}{\rho}+\frac{F_xF_y}{\rho^2}\right]\widehat{\overline{\kappa}}_{xy}. \label{eq:sourceterm8}
\end{eqnarray}
In Eqs.~(\ref{eq:sourceterm6})-(\ref{eq:sourceterm8}), $\widehat{\overline{\kappa}}_{xx}$, $\widehat{\overline{\kappa}}_{yy}$,
$\widehat{\overline{\kappa}}_{xy}$, $\widehat{\overline{\kappa}}_{xyy}$ and $\widehat{\overline{\kappa}}_{xxy}$ can be obtained
from Eqs.~(\ref{eq:kxxtransform})-(\ref{eq:kxxytransform}), respectively.
This can be written in matrix form as
\begin{eqnarray}
\mathcal{K}^T\mathbf{S}=(\mathcal{K}\cdot\mathbf{S})_{\alpha}&=&(\braket{K_0|S_{\alpha}},\braket{K_1|S_{\alpha}},\braket{K_2|S_{\alpha}},\ldots,\braket{K_8|S_{\alpha}}) \nonumber \\
&=& (\widehat{m}^{s}_{0},\widehat{m}^{s}_{1},\widehat{m}^{s}_{2},\ldots,\widehat{m}^{s}_{8})^T\equiv \ket{\widehat{m}^{s}_{\alpha}}. \label{eq:sourceformulation1}
\end{eqnarray}
Now, by exploiting the orthogonal property of $\mathcal{K}$~\cite{premnath09d}, i.e.
$\mathcal{K}^{-1}=\mathcal{K}^T \cdot D^{-1}$, where the diagonal matrix is
$D=\mbox{diag}(\braket{K_0|K_0},\braket{K_1|K_1},\braket{K_2|K_2},\ldots,\braket{K_8|K_8})=\mbox{diag}(9,6,6,36,4,4,12,12,36)$,
we exactly invert Eq.~(\ref{eq:sourceformulation1}) to finally obtain source terms in velocity space $S_{\alpha}$ as
\begin{eqnarray}
S_0&=&\frac{1}{9}\left[-\widehat{m}^{s}_{3}+\widehat{m}^{s}_{8}\right], \label{eq:vsourceterms0}\\
S_1&=&\frac{1}{36}\left[6\widehat{m}^{s}_{1}-\widehat{m}^{s}_{3}+9\widehat{m}^{s}_{4}+6\widehat{m}^{s}_{7}-2\widehat{m}^{s}_{8}\right], \label{eq:vsourceterms1}\\
S_2&=&\frac{1}{36}\left[6\widehat{m}^{s}_{2}-\widehat{m}^{s}_{3}-9\widehat{m}^{s}_{4}+6\widehat{m}^{s}_{6}-2\widehat{m}^{s}_{8}\right], \label{eq:vsourceterms2}\\
S_3&=&\frac{1}{36}\left[-6\widehat{m}^{s}_{1}-\widehat{m}^{s}_{3}+9\widehat{m}^{s}_{4}-6\widehat{m}^{s}_{7}-2\widehat{m}^{s}_{8}\right], \label{eq:vsourceterms3}\\
S_4&=&\frac{1}{36}\left[-6\widehat{m}^{s}_{2}-\widehat{m}^{s}_{3}-9\widehat{m}^{s}_{4}-6\widehat{m}^{s}_{6}-2\widehat{m}^{s}_{8}\right], \label{eq:vsourceterms4}\\
S_5&=&\frac{1}{36}\left[6\widehat{m}^{s}_{1}+6\widehat{m}^{s}_{2}+2\widehat{m}^{s}_{3}
+9\widehat{m}^{s}_{5}-3\widehat{m}^{s}_{6}-3\widehat{m}^{s}_{7}+\widehat{m}^{s}_{8}\right], \label{eq:vsourceterms5}\\
S_6&=&\frac{1}{36}\left[-6\widehat{m}^{s}_{1}+6\widehat{m}^{s}_{2}+2\widehat{m}^{s}_{3}
-9\widehat{m}^{s}_{5}-3\widehat{m}^{s}_{6}+3\widehat{m}^{s}_{7}+\widehat{m}^{s}_{8}\right], \label{eq:vsourceterms6}\\
S_7&=&\frac{1}{36}\left[-6\widehat{m}^{s}_{1}-6\widehat{m}^{s}_{2}+2\widehat{m}^{s}_{3}
+9\widehat{m}^{s}_{5}+3\widehat{m}^{s}_{6}+3\widehat{m}^{s}_{7}+\widehat{m}^{s}_{8}\right], \label{eq:vsourceterms7}\\
S_8&=&\frac{1}{36}\left[6\widehat{m}^{s}_{1}-6\widehat{m}^{s}_{2}+2\widehat{m}^{s}_{3}
-9\widehat{m}^{s}_{5}+3\widehat{m}^{s}_{6}-3\widehat{m}^{s}_{7}+\widehat{m}^{s}_{8}\right]. \label{eq:vsourceterms8}
\end{eqnarray}
This is the explicit set of expressions for the source terms in velocity space $S_{\alpha}$ given in terms of $\overrightarrow{F}$,
$\overrightarrow{u}$ and $\widehat{\overline{\kappa}}_{x^my^n}$, with $2\leq (m+n)\leq 3$ and $0\leq m,n \leq 2$.

Again, using the orthogonal property of $\mathcal{K}$, we can obtain the raw moments of the collision kernel
\begin{equation}
\sum_{\alpha}(\mathcal{K}\cdot \mathbf{\widehat{g}})_{\alpha}e_{\alpha x}^m e_{\alpha y}^n = \sum_{\beta} \braket{K_{\beta}|e_{\alpha x}^m e_{\alpha y}^n}\widehat{g}_{\beta},
\end{equation}
which is of central importance in the subsequent derivation. Note that for collision invariants,
$\widehat{g}_{0}=\widehat{g}_{1}=\widehat{g}_{2}=0$. We get
\begin{eqnarray}
\sum_{\alpha}(\mathcal{K}\cdot \mathbf{\widehat{g}})_{\alpha}=
\sum_{\beta} \braket{K_{\beta}|\rho}\widehat{g}_{\beta}&=&0, \nonumber\\
\sum_{\alpha}(\mathcal{K}\cdot \mathbf{\widehat{g}})_{\alpha}e_{\alpha x}=
\sum_{\beta} \braket{K_{\beta}|e_{\alpha x}}\widehat{g}_{\beta}&=&0, \nonumber\\
\sum_{\alpha}(\mathcal{K}\cdot \mathbf{\widehat{g}})_{\alpha}e_{\alpha y}=
\sum_{\beta} \braket{K_{\beta}|e_{\alpha y}}\widehat{g}_{\beta}&=&0, \nonumber\\
\sum_{\alpha}(\mathcal{K}\cdot \mathbf{\widehat{g}})_{\alpha}e_{\alpha x}^2=
\sum_{\beta} \braket{K_{\beta}|e_{\alpha x}^2}\widehat{g}_{\beta}&=&6\widehat{g}_{3}+2\widehat{g}_{4}, \nonumber\\
\sum_{\alpha}(\mathcal{K}\cdot \mathbf{\widehat{g}})_{\alpha}e_{\alpha y}^2=
\sum_{\beta} \braket{K_{\beta}|e_{\alpha y}^2}\widehat{g}_{\beta}&=&6\widehat{g}_{3}-2\widehat{g}_{4}, \label{eq:collisionkernelmoment}\\
\sum_{\alpha}(\mathcal{K}\cdot \mathbf{\widehat{g}})_{\alpha}e_{\alpha x}e_{\alpha y}=
\sum_{\beta} \braket{K_{\beta}|e_{\alpha x}e_{\alpha y}}\widehat{g}_{\beta}&=&4\widehat{g}_{5}, \nonumber\\
\sum_{\alpha}(\mathcal{K}\cdot \mathbf{\widehat{g}})_{\alpha}e_{\alpha x}^2e_{\alpha y}=
\sum_{\beta} \braket{K_{\beta}|e_{\alpha x}^2e_{\alpha y}}\widehat{g}_{\beta}&=&-4\widehat{g}_{6}, \nonumber\\
\sum_{\alpha}(\mathcal{K}\cdot \mathbf{\widehat{g}})_{\alpha}e_{\alpha x}e_{\alpha y}^2=
\sum_{\beta} \braket{K_{\beta}|e_{\alpha x}e_{\alpha y}^2}\widehat{g}_{\beta}&=&-4\widehat{g}_{7}, \nonumber\\
\sum_{\alpha}(\mathcal{K}\cdot \mathbf{\widehat{g}})_{\alpha}e_{\alpha x}^2e_{\alpha y}^2=
\sum_{\beta} \braket{K_{\beta}|e_{\alpha x}^2e_{\alpha y}^2}\widehat{g}_{\beta}&=&8\widehat{g}_{3}+4\widehat{g}_{8}. \nonumber
\end{eqnarray}

Finally, the LBE in Eq.~(\ref{eq:cascadedLBE3}) can be rewritten in terms of collision and streaming steps, respectively, as
\begin{eqnarray}
\widetilde{\overline{f}}_{\alpha}(\overrightarrow{x},t)&=&\overline{f}_{\alpha}(\overrightarrow{x},t)+\Omega_{{\alpha}(\overrightarrow{x},t)}^{c}+
S_{{\alpha}(\overrightarrow{x},t)},\label{eq:cascadedcollision1}\\
\overline{f}_{\alpha}(\overrightarrow{x}+\overrightarrow{e}_{\alpha},t+1)&=&\widetilde{\overline{f}}_{\alpha}(\overrightarrow{x},t), \label{eq:cascadedstreaming1}
\end{eqnarray}
where the symbol ``tilde" ($\sim$) in the first equation refers to the post-collision state. In terms of the transformed distribution, the hydrodynamic fields can be computed by means of the following:
\begin{eqnarray}
\rho&=&\sum_{\alpha=0}^{8}\overline{f}_{\alpha}=\braket{\overline{f}_{\alpha}|\rho},\label{eq:densitycalculation}\\
\rho u_i&=&\sum_{\alpha=0}^{8}\overline{f}_{\alpha} e_{\alpha i}+\frac{1}{2}F_i=\braket{\overline{f}_{\alpha}|e_{\alpha i}}+\frac{1}{2}F_i, \qquad i \in {x,y}. \label{eq:velocitycalculation}
\end{eqnarray}

\section{\label{sec:cascadedcollisionforcing}Structure of the Collision Operator: Cascaded Central Moments}
Let us now arrive at the expressions for the cascaded formulation of the collision operator using central moments in the presence of
forcing terms based on the results obtained in the last few sections. The basic procedure can be stated as follows. Beginning from
the lowest order central moments that are non-collisional invariants (i.e. $\widehat{\overline{\kappa}}_{xx}$ and higher), they are
successively set equal to their local attractors based on the transformed Maxwellians (Eq.~(\ref{eq:centralmomenttransformed})). This step provides tentative expressions for $\widehat{g}_{\alpha}$ based on the equilibrium assumption. We then modify them to allow for relaxation during collision by multiplying them with corresponding relaxation parameters~\cite{geier06}. In this step, given the cascaded nature of the collision (i.e. $\widehat{g}_{\alpha}\equiv\widehat{g}_{\alpha}(\mathbf{f},\widehat{g}_{\beta}), \beta=0,1,2,\ldots,\alpha-1$, or the dependence of higher order terms on those that are lower orders), care needs to be exercised to multiply the relaxation parameters only with those terms that are not yet in post-collision states (i.e. terms not involving
$\widehat{g}_{\beta}, \beta=0,1,2,\ldots,\alpha-1$ for $\widehat{g}_{\alpha}$). Various details involved in this procedure are given in~\cite{premnath09d}. For brevity, here we summarize the final results which are as follows:
\begin{eqnarray}
\widehat{g}_3&=&\frac{\omega_3}{12}\left\{ \frac{2}{3}\rho+\rho(u_x^2+u_y^2)
-(\widehat{\overline{\kappa}}_{xx}^{'}+\widehat{\overline{\kappa}}_{yy}^{'})
-\frac{1}{2}(\widehat{\sigma}_{xx}^{'}+\widehat{\sigma}_{yy}^{'})
\right\}, \label{eq:collisionkernelg3}\\
\widehat{g}_4&=&\frac{\omega_4}{4}\left\{\rho(u_x^2-u_y^2)
-(\widehat{\overline{\kappa}}_{xx}^{'}-\widehat{\overline{\kappa}}_{yy}^{'})
-\frac{1}{2}(\widehat{\sigma}_{xx}^{'}-\widehat{\sigma}_{yy}^{'})
\right\}, \label{eq:collisionkernelg4}\\
\widehat{g}_5&=&\frac{\omega_5}{4}\left\{\rho u_x u_y
-\widehat{\overline{\kappa}}_{xy}^{'}
-\frac{1}{2}\widehat{\sigma}_{xy}^{'}
\right\}, \label{eq:collisionkernelg5}\\
\widehat{g}_6&=&\frac{\omega_6}{4}\left\{2\rho u_x^2 u_y+\widehat{\overline{\kappa}}_{xxy}^{'}
-2u_x\widehat{\overline{\kappa}}_{xy}^{'}-u_y\widehat{\overline{\kappa}}_{xx}^{'}-\frac{1}{2}\widehat{\sigma}_{xxy}
\right\}-\frac{1}{2}u_y(3\widehat{g}_3+\widehat{g}_4)\nonumber\\&&-2u_x\widehat{g}_5, \label{eq:collisionkernelg6}\\
\widehat{g}_7&=&\frac{\omega_7}{4}\left\{2\rho u_x u_y^2+\widehat{\overline{\kappa}}_{xyy}^{'}
-2u_y\widehat{\overline{\kappa}}_{xy}^{'}-u_x\widehat{\overline{\kappa}}_{yy}^{'}-\frac{1}{2}\widehat{\sigma}_{xyy}
\right\}-\frac{1}{2}u_x(3\widehat{g}_3-\widehat{g}_4)\nonumber\\&&-2u_y\widehat{g}_5, \label{eq:collisionkernelg7}\\
\widehat{g}_8&=&\frac{\omega_8}{4}\left\{\frac{1}{9}\rho+3\rho u_x^2 u_y^2-\left[\widehat{\overline{\kappa}}_{xxyy}^{'}
-2u_x\widehat{\overline{\kappa}}_{xyy}^{'}-2u_y\widehat{\overline{\kappa}}_{xxy}^{'}
+u_x^2\widehat{\overline{\kappa}}_{yy}^{'}+u_y^2\widehat{\overline{\kappa}}_{xx}^{'}\right.\right.
\nonumber \\
&&\left.\left.+4u_xu_y\widehat{\overline{\kappa}}_{xy}^{'}
\right]-\frac{1}{2}\widehat{\sigma}_{xxyy}^{'}
\right\}-2\widehat{g}_3-\frac{1}{2}u_y^2(3\widehat{g}_3+\widehat{g}_4)
-\frac{1}{2}u_x^2(3\widehat{g}_3-\widehat{g}_4)\nonumber\\
&&-4u_xu_y\widehat{g}_5-2u_y\widehat{g}_6
-2u_x\widehat{g}_7.\label{eq:collisionkernelg8}
\end{eqnarray}
In the above, the raw moments of various orders, i.e. $\widehat{\overline{\kappa}}_{x^m y^n}^{'}$ for different $m$ and $n$ are required, which may be obtained from Eqs.~(\ref{eq:rawmomenttransformed0})-(\ref{eq:rawmomenttransformed8}). Similarly, the raw moments of sources of various orders, i.e. $\widehat{\sigma}_{x^m y^n}^{'}$ needed in the above are given in Eqs.~(\ref{eq:rawmomentsourceterm0})-(\ref{eq:rawmomentsourceterm8}) (see Sec.~\ref{sec:rawmoments}). Here, $\omega_{\beta}$, where $\beta=3,4,5,\ldots, 8$, are the relaxation parameters, satisfying $0<\omega_{\beta}<2$. When a multiscale Chapman-Enskog expansion~\cite{chapman64} is applied to this central moment LBM based on central moments, it recovers the Navier-Stokes equation
with the relaxation parameters $\omega_3=\omega^\chi$~and~$\omega_4=\omega_5=\omega^\nu$ controlling bulk and shear viscosities,
respectively (e.g., $\nu=c_s^2\left(\frac{1}{\omega^\nu}-\frac{1}{2}\right)$)~\cite{premnath09d}. The rest of the parameters can
be adjusted to improve numerical stability.

\section{\label{sec:computationalprocedure}Cascaded Central Moment Lattice Boltzmann Equation}
The collision step with the addition of forcing terms (see Eq.~(\ref{eq:cascadecollision1}) and Eq.~(\ref{eq:cascadedcollision1}))
in the stream-and-collide procedure of the LBM, obtained by matching those of the continuous Boltzmann equation as discussed in
the previous sections, is expanded element-wise and can be summarized as follows:
\begin{eqnarray}
\widetilde{\overline{f}}_{0}&=&\overline{f}_{0}+\left[\widehat{g}_0-4(\widehat{g}_3-\widehat{g}_8)\right]+S_0, \nonumber\\
\widetilde{\overline{f}}_{1}&=&\overline{f}_{1}+\left[\widehat{g}_0+\widehat{g}_1-\widehat{g}_3+\widehat{g}_4    +2(\widehat{g}_7-\widehat{g}_8)\right]+S_1, \nonumber\\
\widetilde{\overline{f}}_{2}&=&\overline{f}_{2}+\left[\widehat{g}_0+\widehat{g}_2-\widehat{g}_3-\widehat{g}_4
+2(\widehat{g}_6-\widehat{g}_8)\right]+S_2, \nonumber\\
\widetilde{\overline{f}}_{3}&=&\overline{f}_{3}+\left[\widehat{g}_0-\widehat{g}_1-\widehat{g}_3+\widehat{g}_4
-2(\widehat{g}_7+\widehat{g}_8)\right]+S_3, \nonumber\\
\widetilde{\overline{f}}_{4}&=&\overline{f}_{4}+\left[\widehat{g}_0-\widehat{g}_2-\widehat{g}_3-\widehat{g}_4
-2(\widehat{g}_6+\widehat{g}_8)\right]+S_4, \\
\widetilde{\overline{f}}_{5}&=&\overline{f}_{5}+\left[\widehat{g}_0+\widehat{g}_1+\widehat{g}_2+2\widehat{g}_3
+\widehat{g}_5-\widehat{g}_6-\widehat{g}_7+\widehat{g}_8\right]+S_5, \nonumber\\
\widetilde{\overline{f}}_{6}&=&\overline{f}_{6}+\left[\widehat{g}_0-\widehat{g}_1+\widehat{g}_2+2\widehat{g}_3
-\widehat{g}_5-\widehat{g}_6+\widehat{g}_7+\widehat{g}_8\right]+S_6, \nonumber\\
\widetilde{\overline{f}}_{7}&=&\overline{f}_{7}+\left[\widehat{g}_0-\widehat{g}_1-\widehat{g}_2+2\widehat{g}_3
+\widehat{g}_5+\widehat{g}_6+\widehat{g}_7+\widehat{g}_8\right]+S_7, \nonumber\\
\widetilde{\overline{f}}_{8}&=&\overline{f}_{8}+\left[\widehat{g}_0+\widehat{g}_1-\widehat{g}_2+2\widehat{g}_3
-\widehat{g}_5+\widehat{g}_6-\widehat{g}_7+\widehat{g}_8\right]+S_8. \nonumber
\end{eqnarray}
The collision kernel $\widehat{g}_{\beta}$ needed here can be computed from the expressions given in the previous section (see Sec.~\ref{sec:cascadedcollisionforcing}). The source terms in the velocity space can be obtained from Eqs.~(\ref{eq:vsourceterms0})-(\ref{eq:vsourceterms8}) (see Sec.~\ref{sec:rawmoments}). The remaining streaming step is carried out as usual by using the post collision values $\widetilde{\overline{f}}_{\alpha}$ obtained from above. Once the local distribution function is known, macroscopic fluid density and velocity fields satisfying the Galilean invariant Navier-Stokes equations in the presence of force fields can be obtained from Eqs.~(\ref{eq:densitycalculation}) and (\ref{eq:velocitycalculation}), respectively.

\section{\label{sec:cascadedcollisionreducedcompressibility}Cascaded Collision Operator with Reduced Compressibility Effects}
While a main goal of this work is the introduction of a self-consistent approach based on the continuous Boltzmann equation to incorporate non-equilibrium effects into the central moment approach for general applicability, it is also useful to consider its limiting cases. For example, the incompressible limit of fluid flow corresponds to considering very small deviations from the local equilibrium, a special case with various applications. In particular, this would allow simple representation of incompressible turbulence considered later in this work.

Being a kinetic approach, the lattice Boltzmann method is inherently compressible in nature. On the other hand, when it is
desired to reproduce the ``incompressible'' Navier-Stokes equations as mentioned above, it is important to reduce such compressibility effects. An approach in this regard was introduced earlier in~\cite{he97a}. Here, we will extend it further in the context of the central
moment LBM in the presence of forcing. It may be noted that the fundamental expressions for the continuous central moments for the local equilibrium as well as the forcing given in Secs.~\ref{sec:ccentralmomentsfandfeq} and \ref{sec:ccentralmomentssources}, respectively, from which their discrete counterparts are derived, remains unchanged for this case. However, the key element to incorporate a systematic reduction of compressibility effects lies in the following careful definition of the raw moments of the
hydrodynamic fields:
\begin{eqnarray}
\rho&=&\sum_{\alpha=0}^{8}f_{\alpha}=\sum_{\alpha=0}^{8}\overline{f}_{\alpha},\\
\rho_0\overrightarrow{u}&=&\sum_{\alpha=0}^{8}f_{\alpha}\overrightarrow{e}_{\alpha}=\sum_{\alpha=0}^{8}\overline{f}_{\alpha}\overrightarrow{e}_{\alpha}+\frac{1}{2}\overrightarrow{F}.
\end{eqnarray}
where $\rho=\rho_0+\delta \rho$. Here $\rho_0$ and $\delta \rho$ are the constant reference value and fluctuations of density,
respectively. That is, in the above, contributions of density fluctuations are eliminated from first-order moments representing the components of momentum. Thus, we get
\begin{eqnarray}
\braket{\widetilde{\overline{f}}_{\alpha}|\rho}&=&\rho,\\
\braket{\widetilde{\overline{f}}_{\alpha}|e_{\alpha x}}&=&\rho_0u_x+\frac{1}{2}F_x,\\
\braket{\widetilde{\overline{f}}_{\alpha}|e_{\alpha y}}&=&\rho_0u_y+\frac{1}{2}F_y.
\end{eqnarray}
Using the procedure discussed in the previous sections and with the above specialized re-definition of the conserved moments, we
obtain, after some simplification, the cascaded collision operator with reduced compressibility effects. They are reported here in
the following:
\begin{eqnarray}
\widehat{g}_3&=&\frac{\omega_3}{12}\left\{ \frac{2}{3}\rho+(\rho_0-\delta \rho)(u_x^2+u_y^2)
-(\widehat{\overline{\kappa}}_{xx}^{'}+\widehat{\overline{\kappa}}_{yy}^{'})
-\frac{1}{2}(\widehat{\sigma}_{xx}^{'}+\widehat{\sigma}_{yy}^{'})
\right\}, \label{eq:collisionkernelcompact3}\\
\widehat{g}_4&=&\frac{\omega_4}{4}\left\{(\rho_0-\delta \rho)(u_x^2-u_y^2)
-(\widehat{\overline{\kappa}}_{xx}^{'}-\widehat{\overline{\kappa}}_{yy}^{'})
-\frac{1}{2}(\widehat{\sigma}_{xx}^{'}-\widehat{\sigma}_{yy}^{'})
\right\}, \label{eq:collisionkernelcompact4}\\
\widehat{g}_5&=&\frac{\omega_5}{4}\left\{(\rho_0-\delta \rho) u_x u_y
-\widehat{\overline{\kappa}}_{xy}^{'}
-\frac{1}{2}\widehat{\sigma}_{xy}^{'}
\right\}, \label{eq:collisionkernelcompact5}\\
\widehat{g}_6&=&\frac{\omega_6}{4}\left\{(2\rho_0-\delta \rho) u_x^2 u_y+\widehat{\overline{\kappa}}_{xxy}^{'}
-2u_x\widehat{\overline{\kappa}}_{xy}^{'}-u_y\widehat{\overline{\kappa}}_{xx}^{'}-\frac{1}{2}\widehat{\sigma}_{xxy}
\right\}-\frac{1}{2}u_y(3\widehat{g}_3+\widehat{g}_4)\nonumber\\&&-2u_x\widehat{g}_5, \label{eq:collisionkernelcompact6}\\
\widehat{g}_7&=&\frac{\omega_7}{4}\left\{(2\rho_0-\delta \rho) u_x u_y^2+\widehat{\overline{\kappa}}_{xyy}^{'}
-2u_y\widehat{\overline{\kappa}}_{xy}^{'}-u_x\widehat{\overline{\kappa}}_{yy}^{'}-\frac{1}{2}\widehat{\sigma}_{xyy}
\right\}-\frac{1}{2}u_x(3\widehat{g}_3-\widehat{g}_4)\nonumber\\&&-2u_y\widehat{g}_5, \label{eq:collisionkernelcompact7}\\
\widehat{g}_8&=&\frac{\omega_8}{4}\left\{\frac{1}{9}\rho+(3\rho_0-\delta \rho) u_x^2 u_y^2-\left[\widehat{\overline{\kappa}}_{xxyy}^{'}
-2u_x\widehat{\overline{\kappa}}_{xyy}^{'}-2u_y\widehat{\overline{\kappa}}_{xxy}^{'}
+u_x^2\widehat{\overline{\kappa}}_{yy}^{'}+u_y^2\widehat{\overline{\kappa}}_{xx}^{'}\right.\right.
\nonumber \\
&&\left.\left.+4u_xu_y\widehat{\overline{\kappa}}_{xy}^{'}
\right]-\frac{1}{2}\widehat{\sigma}_{xxyy}^{'}
\right\}-2\widehat{g}_3-\frac{1}{2}u_y^2(3\widehat{g}_3+\widehat{g}_4)
-\frac{1}{2}u_x^2(3\widehat{g}_3-\widehat{g}_4)\nonumber\\
&&-4u_xu_y\widehat{g}_5-2u_y\widehat{g}_6
-2u_x\widehat{g}_7.\label{eq:collisionkernelcompact8}
\end{eqnarray}
The above collision operator that selectively introduces density fluctuations where necessary can reduce compressibility effects for a inertial frame invariant flow field while retaining its linear acoustics. It can thus allow for a better comparison with ``incompressible'' macroscopic fluid dynamic equations, particularly for turbulent flows as discussed later.

\section{\label{sec:factorizedcentralmomentmethod}Factorized Central Moment Model for Collision}
In this section, we will derive an alternative form of the central moment LBE with forcing terms based on a different choice
of the local attractor in the collision step for improved physical coherence. Continuous kinetic theory, as originally initiated by Maxwell~\cite{maxwell60}, features two important properties for the local equilibrium or the Maxwell distribution -- Galilean invariance and factorization in Cartesian components of the particle velocity. As discussed recently~\cite{geier09,karlin10}, it could prove useful to inherent these properties at the discrete particle velocity level. The use of central moments maintains Galilean invariance by construction. Factorization property of the distribution function implies that particle velocities are random variables. An extension of the factorization idea beyond equilibrium as a model for describing the relaxation process during collision
was proposed to construct local attractors~\cite{geier09}. Specifically, the basic postulate behind this model is that the
Cartesian products of the post-collision values of the orthogonal central moments of lower orders that are not in equilibrium
forms as the basis for the attractors of the higher order moments. Here, we further extend this to include source terms
so that the model can incorporate force fields. Thus, the attractors for central moments of different orders are given as \begin{eqnarray}
\widehat{\kappa}_{x}^{at}&=&\widetilde{\widehat{\kappa}}_x=0,\label{eq:factorx}\\
\widehat{\kappa}_{y}^{at}&=&\widetilde{\widehat{\kappa}}_y=0,\label{eq:factory}\\
\widehat{\kappa}_{xy}^{at}&=&\widetilde{\widehat{\kappa}}_x \widetilde{\widehat{\kappa}}_y=0,\label{eq:factorxy}\\
\widehat{\kappa}_{xxy}^{at}&=&\widetilde{\widehat{\kappa}}_{xx} \widetilde{\widehat{\kappa}}_y=0,\label{eq:factorxxy}\\
\widehat{\kappa}_{xyy}^{at}&=&\widetilde{\widehat{\kappa}}_x \widetilde{\widehat{\kappa}}_{yy}=0,\label{eq:factorxyy}\\
\widehat{\kappa}_{xxyy}^{at}&=&\widetilde{\widehat{\kappa}}_{xx} \widetilde{\widehat{\kappa}}_{yy},\label{eq:factorxxyy}
\end{eqnarray}
while the second-order longitudinal central moments are obtained from Maxwellian as given in an earlier section
(see Sec.~\ref{sec:discretecentralmoments}), i.e. $\widehat{\kappa}_{xx}^{at}=\widehat{\kappa}_{yy}^{at}=\rho c_s^2$. In essence,
the distinguishing feature of the factorized central moment lies in the use of modified attractors for third and higher order moments.
Now, using the following central moment identity of the post-collision state
$\widetilde{\widehat{\kappa}}_{x^my^n}=\widehat{\kappa}_{x^my^n}+\sum_{\beta}\braket{K_{\alpha}|(e_{\alpha x}-u_x)^m(e_{\alpha y}-u_y)^n}\widehat{g}_{\beta}$,
for $m=2,n=0$ and $m=0,n=2$, we get
\begin{eqnarray}
\widetilde{\widehat{\kappa}}_{xx}&=&\widehat{\kappa}_{xx}+(6\widehat{g}_3+2\widehat{g}_4),\\
\widetilde{\widehat{\kappa}}_{yy}&=&\widehat{\kappa}_{yy}+(6\widehat{g}_3-2\widehat{g}_4).
\end{eqnarray}
Note that it also follows that
$\widehat{\kappa}_{xx}=\widehat{\overline{\kappa}}_{xx}$ and $\widehat{\kappa}_{yy}=\widehat{\overline{\kappa}}_{yy}$.
We can then rewrite everything in terms of transformed raw moments, i.e.
$\widehat{\overline{\kappa}}_{xx}=\widehat{\overline{\kappa}}_{xx}^{'}-\rho u_x^2+F_xu_x$ and
$\widehat{\overline{\kappa}}_{yy}=\widehat{\overline{\kappa}}_{yy}^{'}-\rho u_y^2+F_yu_y$. These yield \begin{eqnarray}
\widetilde{\widehat{\kappa}}_{xx}&=&\widehat{\overline{\kappa}}_{xx}^{'}-\rho u_x^2+F_xu_x+(6\widehat{g}_3+2\widehat{g}_4),\\
\widetilde{\widehat{\kappa}}_{yy}&=&\widehat{\overline{\kappa}}_{yy}^{'}-\rho u_y^2+F_yu_y+(6\widehat{g}_3-2\widehat{g}_4).
\end{eqnarray}
In effect, the attractor for the fourth-order moment, i.e. Eq.~(\ref{eq:factorxxyy}) reduces to
\begin{eqnarray}
\widehat{\kappa}_{xxyy}^{at}&=&\left[\widehat{\overline{\kappa}}_{xx}^{'}-\rho u_x^2+F_xu_x+(6\widehat{g}_3+2\widehat{g}_4)\right]\times \nonumber\\
&&\left[\widehat{\overline{\kappa}}_{yy}^{'}-\rho u_y^2+F_yu_y+(6\widehat{g}_3-2\widehat{g}_4)\right].
\end{eqnarray}

Now, to obtain an operational step in terms of the transformed variables, we use the relation
$\widehat{\overline{\kappa}}_{x^my^n}^{at}=\widehat{\kappa}_{x^my^n}^{at}-\frac{1}{2}\widehat{\sigma}_{x^my^n}$ to finally
get the following expression for the fourth-order central moment
\begin{eqnarray}
\widehat{\overline{\kappa}}_{xxyy}^{at}&=&\left[\widehat{\overline{\kappa}}_{xx}^{'}-\rho u_x^2+F_xu_x+(6\widehat{g}_3+2\widehat{g}_4)\right]
\left[\widehat{\overline{\kappa}}_{yy}^{'}-\rho u_y^2+F_yu_y+(6\widehat{g}_3-2\widehat{g}_4)\right]\nonumber\\
&&-\frac{F_x}{\rho}\widehat{\overline{\kappa}}_{xyy}-
\frac{F_y}{\rho}\widehat{\overline{\kappa}}_{xxy}-\frac{F_y^2}{2\rho^2}\widehat{\overline{\kappa}}_{xx}-
\frac{F_x^2}{2\rho^2}\widehat{\overline{\kappa}}_{yy}-2\frac{F_xF_y}{\rho^2}\widehat{\overline{\kappa}}_{xy}. \label{eq:factorxxyyfinal}
\end{eqnarray}
By replacing $\widehat{\overline{\kappa}}_{x^my^n}^{\mathcal{M}}$ with $\widehat{\overline{\kappa}}_{x^my^n}^{at}$ as given above,
we can derive the collision kernel with factorized central moments as attractors. It follows that the expressions in
Sec.~\ref{sec:cascadedcollisionforcing} for $\widehat{g}_{\beta}$, $\beta=3,4,5,6,7$ are the same as before with the
exception for $\widehat{g}_{8}$. The expression for $\widehat{g}_{8}$ in Eq.~(\ref{eq:collisionkernelg8}) is modified such
that term $\frac{1}{9}\rho(=\widehat{\overline{\kappa}}_{xxyy}^{\mathcal{M}})$ in this equation is now replaced by
$\widehat{\overline{\kappa}}_{xxyy}^{at}$ given in Eq.~(\ref{eq:factorxxyyfinal}).
In a similar vein, the above expression can be modified for reduced compressibility effects
(see Sec.~\ref{sec:cascadedcollisionreducedcompressibility}) as
\begin{eqnarray}
\widehat{\kappa}_{xxyy}^{at}&=&\left[\widehat{\overline{\kappa}}_{xx}^{'}-(\rho_0-\delta \rho) u_x^2+F_xu_x+(6\widehat{g}_3+2\widehat{g}_4)\right]\times \nonumber\\
&&\left[\widehat{\overline{\kappa}}_{yy}^{'}-(\rho_0-\delta \rho) u_y^2+F_yu_y+(6\widehat{g}_3-2\widehat{g}_4)\right]\nonumber\\
&&-\frac{F_x}{\rho}\widehat{\overline{\kappa}}_{xyy}-
\frac{F_y}{\rho}\widehat{\overline{\kappa}}_{xxy}-\frac{F_y^2}{2\rho^2}\widehat{\overline{\kappa}}_{xx}-
\frac{F_x^2}{2\rho^2}\widehat{\overline{\kappa}}_{yy}-2\frac{F_xF_y}{\rho^2}\widehat{\overline{\kappa}}_{xy} \end{eqnarray}
to modify $\widehat{g}_{8}$ in Eq.~(\ref{eq:collisionkernelcompact8}).

\section{\label{sec:Galileaninvariantturbmodels}Galilean Invariant Filtered Turbulence Representation using Lattice Kinetic Framework}
Based on the various elements derived in the previous sections, we are now in a position to construct an approach for simulation of
Galilean invariant turbulent flow field by incorporating appropriate turbulence models in the LBM. The starting point in the statistical continuum description of turbulence is the Reynolds decomposition of the velocity field of the fluid into `resolved' and `unresolved' parts. The resolved part is obtained by applying either some averaging in space or time (in the Reynolds Averaged Navier-Stokes (RANS) context) or by applying a filter (in the LES). Application of this decomposition to the Navier-Stokes (NS) equation leads to additional unknown terms involving products of the unresolved fields, which are Reynolds stresses (in RANS) or the subgrid stresses (in LES). This closure problem then becomes the main focus of turbulence modeling. Due to the scale invariance property of the NS equations~\cite{germano92}, the averaged and the filtered equations, as well as the additional stress-like closure terms have similar forms. Thus, a unified statistical approach may be adopted for turbulence modeling. It is interesting to note that ideas based on kinetic theory provided the original inspiration for the Reynolds decomposition~\cite{reynolds95} as well as early works on developing turbulence models.

The underlying motivation here is to develop a unified statistical averaged description (for RANS) or formal spatial filtered
representation (for LES) of \emph{inertial frame invariant} turbulence in a kinetic approach based on the LBM derived in earlier sections. This would also allow reconciliation of continuum and non-continuum effects on turbulence as discussed recently~\cite{girimaji07}. The following notation for Reynolds decomposition is adopted here. For any scalar $\phi$, vector $\overrightarrow{v}$ and tensor $T_{ij}$, we have
\begin{eqnarray}
\phi&=&\underline{\phi}+\phi^{'}, \quad \mbox{with} \quad \underline{\phi^{'}}=0,\nonumber\\
\overrightarrow{v}&=&\underline{\overrightarrow{v}}+\overrightarrow{v}^{'}, \quad \mbox{with} \quad \underline{\overrightarrow{v}^{'}}=0,\nonumber\\
T_{ij}&=&\underline{T_{ij}}+T_{ij}^{'}, \quad \mbox{with} \quad \underline{T_{ij}^{'}}=0,\nonumber
\end{eqnarray}
where $\underline{(\cdot)}$ is an operator representing either some form of statistical average or filter to obtain the resolved part
and the symbols with primes denote the unresolved parts of the field. As discussed in~\cite{girimaji07}, application of the above decomposition directly to the continuous Boltzmann equation
(Eq.~(\ref{eq:continuousBoltzmannequation})) leads to certain difficulties. In particular, using $f=\underline{f}+f^{'}$
for the distribution function in Eq.~(\ref{eq:continuousBoltzmannequation}), which leads to a statistically averaged
kinetic equation, does not provide a clear interpretation of turbulence physics. The local collision term needs to model
all essential physics, including the non-linear and non-local momentum transfer effects of turbulence. Moreover, the use
of the averaged attractor based on the Maxwellian
$\underline{f^{\mathcal{M}}}$ within the collision term $\underline{\Omega(f,f)}$ leads to modeling difficulties
since
$\underline{\exp\left[-\frac{(\overrightarrow{\xi}-\overrightarrow{u})^2}{2c_s^2}\right]}\neq \exp\left[-\frac{(\overrightarrow{\xi}-\underline{\overrightarrow{u}})^2}{2c_s^2}\right]$.
Thus, an alternative approach is needed to coherently represent continuum/kinetic effects on turbulence.

To circumvent these issues, a transformation for the velocities was recently suggested~\cite{girimaji07}, which is adopted here
to provide Galilean invariant turbulence representation. The key element is to clearly separate the advective turbulence effects due
to unresolved velocity field $\overrightarrow{u}^{'}$ from the dissipative collision that represent microscopic effects. This
is accomplished by an inspection of the local Maxwellian given in terms of the microscopic particle velocity $\overrightarrow{\xi}$
and the macroscopic fluid velocity $\overrightarrow{u}$. That is, it consists of the term involving the peculiar velocity \begin{equation}
\overrightarrow{\xi}-\overrightarrow{u}\nonumber
\end{equation}
as its argument which should be made independent of the unresolved part of the macroscopic fluid velocity $\overrightarrow{u}^{'}$,
when the averaging operator is applied. That is,
\begin{equation}
\overrightarrow{\xi}-(\underline{\overrightarrow{u}}+\overrightarrow{u}^{'})=(\overrightarrow{\xi}-\overrightarrow{u}^{'})-\underline{\overrightarrow{u}}\nonumber
\end{equation}
should be transformed appropriately, which can be accomplished by defining a new variable $\overrightarrow{\eta}$ as
\begin{equation}
\overrightarrow{\eta}=\overrightarrow{\xi}-\overrightarrow{u}^{'}.
\end{equation}
Now, the Maxwellian in the transformed peculiar velocity $\overrightarrow{\eta}-\underline{\overrightarrow{u}}$ commutes with the
operator for averaging or filtering. That is,
$
\underline{\exp\left[-\frac{(\overrightarrow{\eta}-\overrightarrow{u})^2}{2c_s^2}\right]}=\exp\left[-\frac{(\overrightarrow{\eta}-\underline{\overrightarrow{u}})^2}{2c_s^2}\right].
$
This facilitates the separation of various aspects of turbulence physics modeling.
Based on this a new distribution function $h(\overrightarrow{x},\overrightarrow{\eta},t)$ and its local Maxwellian are defined by
\begin{equation}
h(\overrightarrow{x},\overrightarrow{\eta},t)=f(\overrightarrow{x},\overrightarrow{\xi},t), \label{eq:modifiedh}
\end{equation}
and
\begin{equation}
h^{\mathcal{M}}(\overrightarrow{\eta},\overrightarrow{u})=\frac{\rho}{2\pi c_s^2}\exp\left[-\frac{(\overrightarrow{\eta}-\underline{\overrightarrow{u}})^2}{2c_s^2}\right],
\end{equation}
respectively. The continuous Boltzmann equation, i.e. Eq.~(\ref{eq:continuousBoltzmannequation}) (without the forcing term for simplicity) is then transformed into a modified kinetic equation in terms of $\eta$ and $h$ as follows. From Eq.~(\ref{eq:modifiedh}),
$\overrightarrow{\nabla}_{\eta}h=\overrightarrow{\nabla}_{\xi}f$. When $\eta=\mbox{constant}$, we have
$(\overrightarrow{\nabla}_{x}\overrightarrow{\xi})_{\eta}=\overrightarrow{\nabla}_{x}\overrightarrow{u}^{'}$ and
$(\partial_t\overrightarrow{\xi})_{\eta}=\partial_t\overrightarrow{u}^{'}$. Hence, the derivatives in new variables are
\begin{eqnarray}
\overrightarrow{\nabla}_{x}h&=&\overrightarrow{\nabla}_{x}f+(\overrightarrow{\nabla}_{x}\overrightarrow{\xi})_{\eta}\cdot\overrightarrow{\nabla}_{\xi}f
=\overrightarrow{\nabla}_{x}f+(\overrightarrow{\nabla}_{x}\overrightarrow{u}^{'})_{\eta}\cdot\overrightarrow{\nabla}_{\eta}h, \nonumber \\
\partial_th&=&\partial_t f+(\partial_t\overrightarrow{\xi})_{\eta}\cdot\overrightarrow{\nabla}_{\xi}f
=\partial_t f+(\partial_t \overrightarrow{u}^{'})_{\eta}\cdot\overrightarrow{\nabla}_{\eta}h. \nonumber
\end{eqnarray}
The continuous Boltzmann equation is thus modified to~\cite{girimaji07}
\begin{equation}
\partial_t h +\overrightarrow{\eta}\cdot\overrightarrow{\nabla}_x h+\overrightarrow{u}^{'}\cdot\overrightarrow{\nabla}_x h-\overrightarrow{a}^{'}\cdot\overrightarrow{\nabla}_{\eta}h=\Omega(h,h), \label{eq:modifiedBoltzmannequation}
\end{equation}
where $\overrightarrow{a}^{'}=\partial_t \overrightarrow{u}^{'}+\eta\cdot\overrightarrow{\nabla}_x\overrightarrow{u}^{'}+
\overrightarrow{u}^{'}\cdot\overrightarrow{\nabla}_x\overrightarrow{u}^{'}$. Considering incompressible flows, where the
unresolved velocity field satisfies $\overrightarrow{\nabla}\cdot\overrightarrow{u}^{'}=0$, we get
$\overrightarrow{u}^{'}\cdot\overrightarrow{\nabla}_x h=\overrightarrow{\nabla}_x (h \overrightarrow{u}^{'})$ and
$\overrightarrow{a}^{'}\cdot\overrightarrow{\nabla}_x h=\overrightarrow{\nabla}_x (h \overrightarrow{a}^{'})$. As a result,
Eq.~(\ref{eq:modifiedBoltzmannequation}) is further simplified to
\begin{equation}
\partial_t h +\overrightarrow{\eta}\cdot\overrightarrow{\nabla}_x h+\overrightarrow{\nabla}_x (h \overrightarrow{u}^{'})-\overrightarrow{\nabla}_x (h \overrightarrow{a}^{'})=\Omega(h,h).\label{eq:modifiedBoltzmannequation1}
\end{equation}
Now, applying the statistical averaging or filtering operator on Eq.~(\ref{eq:modifiedBoltzmannequation1}) and using $\underline{\Omega(h,h)}=\Omega(\underline{h},\underline{h})$, we get the new kinetic equation
\begin{equation}
\partial_t \underline{h} +\overrightarrow{\eta}\cdot\overrightarrow{\nabla}_x \underline{h}+\overrightarrow{\nabla}_x (\underline{h \overrightarrow{u}^{'}})-\overrightarrow{\nabla}_x (\underline{h \overrightarrow{a}^{'}})=\Omega(\underline{h},\underline{h}).\label{eq:modifiedaveragedBoltzmannequation}
\end{equation}
The averaged density and momentum can then be obtained by taking moments of $\overline{h}$. That is,
$
\underline{\rho}= \int \underline{h} d\overrightarrow{\eta},\underline{\rho \overrightarrow{u}}=\int \underline{h}\overrightarrow{\eta} d\overrightarrow{\eta}.
$
Now, $\underline{h(\eta)}$ using Eq.~(\ref{eq:modifiedaveragedBoltzmannequation}) is better suited to represent
turbulence physics for the following reasons. The term involving $\underline{h \overrightarrow{u}^{'}}$ represents transport in
physical space, i.e. redistributes $\underline{h}$ to smoothen any gradients. $\underline{h \overrightarrow{a}^{'}}$ represents
transport in velocity phase space, and acts as a source/sink for energy cascade~\cite{girimaji07}. In particular, both these
quantities can be directly related to continuum based closure models. On the other hand, the role of collision operator is
then to simply represent averaged effect of irreversible molecular collisions. The averaged kinetic equation,
Eq.~(\ref{eq:modifiedaveragedBoltzmannequation}), can be further simplified by considering the following simple microscopic
closure~\cite{girimaji07}
\begin{eqnarray}
\underline{h \overrightarrow{u}^{'}}&\approx&\underline{h}\thinspace\thinspace\underline{\overrightarrow{u}^{'}}=0,\\
\underline{h \overrightarrow{a}^{'}}&\approx&\underline{h}\thinspace\thinspace\underline{\overrightarrow{a}^{'}}=
\underline{h}\overrightarrow{\nabla}_x\cdot(\underline{\overrightarrow{u}^{'}\overrightarrow{u}^{'}})
\end{eqnarray}
That is, $\underline{h}$ is uncorrelated with both $\underline{u}^{'}_i$ and $\underline{a}^{'}_i$, which
reproduces the averaged momentum equations with additional Reynolds stress terms $\underline{u_i^{'}u_j^{'}}$
that can be closed by means of any conventional macroscopic turbulence models. Thus,
Eq.~(\ref{eq:modifiedaveragedBoltzmannequation}) can now be rewritten as
\begin{equation}
\partial_t \underline{h} +\overrightarrow{\eta}\cdot\overrightarrow{\nabla}_x \underline{h}=\Omega(\underline{h},\underline{h})+
\overrightarrow{\nabla}_x\cdot(\underline{\overrightarrow{u}^{'}\overrightarrow{u}^{'}})\cdot \overrightarrow{\nabla}_{\eta}\underline{h},\label{eq:modifiedaveragedBoltzmannequation1}
\end{equation}
which represents the evolution of the statistical averaged/filtered turbulence field by means of the Reynolds stresses that
appear as a forcing term in a kinetic framework.

Let us now develop a Galilean invariant lattice kinetic equation, i.e. which provides inertial frame invariant representation
with respect to the \emph{resolved} velocity field obtained by statistical averaging/filtering.
For brevity and to avoid the use of additional new notations, let us rewrite Eq.~(\ref{eq:modifiedaveragedBoltzmannequation1})
by replacing $\underline{h}$ by $\underline{f}$ (and $\overline{\eta}$ and $\overline{\xi}$) to make use of the developments of
the previous sections. That is,
\begin{equation}
\partial_t \underline{f} +\overrightarrow{\xi}\cdot\overrightarrow{\nabla}_x \underline{f}=\Omega(\underline{f},\underline{f})+
\overrightarrow{\nabla}_x\cdot(\underline{\overrightarrow{u}^{'}\overrightarrow{u}^{'}})\cdot \overrightarrow{\nabla}_{\xi}\underline{f}.
\end{equation}
from which the resolved hydrodynamic fields can be defined as follows:
\begin{equation}
\underline{\rho}=\int \underline{f} d\overrightarrow{\xi},\qquad \underline{\rho \overrightarrow{u}}=\int \underline{f} \overrightarrow{\xi} d\overrightarrow{\xi}.
\end{equation}

Now, we define operator averaged continuous central moments as
\begin{equation}
\underline{\widehat{\Pi}}_{x^my^n}=\int_{-\infty}^{\infty}\int_{-\infty}^{\infty}\underline{f}(\xi_x-\underline{u_x})^m(\xi_y-\underline{u_y})^nd\xi_xd\xi_y
\label{eq:resolvedcentralmomentfdefinition}
\end{equation}
and similarly for the continuous central moments of the Maxwellian $\underline{\widehat{\Pi}}_{x^my^n}^{\mathcal{M}}$ based on
replacing $h^{\mathcal{M}}$ and $\overrightarrow{\eta}$ by $f^{\mathcal{M}}$ and $\overrightarrow{\xi}$, respectively.
The Cartesian components of the unresolved turbulent Reynolds stresses may be written as
\begin{eqnarray}
\underline{a_x^{'}}&=&-\partial_x(\underline{u_x^{'}u_x^{'}})-\partial_y(\underline{u_x^{'}u_y^{'}}),\\
\underline{a_y^{'}}&=&-\partial_x(\underline{u_x^{'}u_y^{'}})-\partial_y(\underline{u_y^{'}u_y^{'}}),
\end{eqnarray}
where $\underline{\overrightarrow{a}^{'}}=(\underline{a_x^{'}}, \underline{a_y^{'}})$, from which we
analogously define a source/sink continuous central moment as
\begin{equation}
\underline{\widehat{\Gamma}^{a}_{x^my^n}}=\int_{-\infty}^{\infty}\int_{-\infty}^{\infty}\underline{\delta f^{a^{'}}}(\xi_x-\underline{u_x})^m(\xi_y-\underline{u_y})^nd\xi_xd\xi_y.
\label{eq:resolvedcentralmomentforce}
\end{equation}
Here, $\underline{\delta f^{a^{'}}}=-\underline{\overrightarrow{a}^{'}}\cdot \overrightarrow{\nabla}_{\xi}\underline{f}$.
It readily follows from Eq.~(\ref{eq:centralmomentforceexactidentity2}) that $\underline{\widehat{\Gamma}}^{a}_{x^my^n}$ also satisfies the following exact identity $\underline{\widehat{\Gamma}}^{a}_{x^my^n}=m a_x^{'}\underline{\widehat{\Pi}}_{x^{m-1}y^n}+n a_y^{'}\underline{\widehat{\Pi}}_{x^my^{n-1}}$.
That is, the statistical averaged/filtered central moment of sources/sinks due to unresolved fields of a given order is dependent on the product of the Cartesian components of the gradients of turbulent stresses with the lower order central moments of
the averaged/filtered distribution function. The corresponding discrete central moment LBM can be devised by considering the
following averaged representation of discrete vectors supported by the particle velocity set:
$\mathbf{\underline{f}}=\ket{\underline{f_{\alpha}}}=(\underline{f}_0,\underline{f}_1,\underline{f}_2,\ldots,\underline{f}_8)^T$,
$\mathbf{\widehat{\underline{g}}}=\ket{\widehat{\underline{g}}_{\alpha}}=(\widehat{\underline{g}}_0,\widehat{\underline{g}}_1,
\widehat{\underline{g}}_2,\ldots,\widehat{\underline{g}}_8)^T$, $\mathbf{\underline{S}}=\ket{\underline{S_{\alpha}}}=(\underline{S}_0,\underline{S}_1,\underline{S}_2,\ldots,\underline{S}_8)^T$, and
$\bm{\underline{\Omega}}^{c}\equiv \bm{\underline{\Omega}}^{c}(\underline{\mathbf{f}},\mathbf{\underline{\widehat{g}}})=
(\mathcal{K}\cdot \mathbf{\underline{\widehat{g}}})=(\underline{\Omega}_0^{c},\underline{\Omega}_1^{c},\underline{\Omega}_2^{c},\ldots,\underline{\Omega}_8^{c})^T$,
and invoking Galilean invariance matching principle, i.e. matching the continuous and discrete central moments of various quantities at successively higher orders as discussed in earlier sections. In particular, the statistical averaged/filtered discrete collision operator $\bm{\underline{\Omega}}^{c}$ can be obtained by considering reduced compressibility effects and factorized attractors as in Sec.~\ref{sec:factorizedcentralmomentmethod}. Furthermore, the corresponding source terms in velocity space $\mathbf{\underline{S}}$ can be constructed using the procedure outlined in Sec.~\ref{sec:rawmoments}. The operator averaged LBE, in terms of the transformed distribution function $\underline{\overline{f}}_{\alpha}$ for improved accuracy, can be finally written as
\begin{equation}
\underline{\overline{f}}_{\alpha}(\overrightarrow{x}+\overrightarrow{e}_{\alpha},t+1)=
\underline{\overline{f}}_{\alpha}(\overrightarrow{x},t)+\underline{\Omega}_{{\alpha}(\overrightarrow{x},t)}^{c}+
\underline{S}_{{\alpha}(\overrightarrow{x},t)},
\label{eq:resolvedcascadedLBE}
\end{equation}
where $\underline{\overline{f}}_{\alpha}=\underline{f}_{\alpha}-\frac{1}{2}\underline{S}_{\alpha}$. Here, as before, we have adopted the standard discretization for the streaming step (see the comment following Eq.~(\ref{eq:cascadedLBE3})). The resolved hydrodynamic fields in the reduced compressibility formulation can then be obtained as
\begin{eqnarray}
\underline{\rho}&=&\sum_{\alpha=0}^{8}\underline{\overline{f}}_{\alpha}=\braket{\underline{\overline{f}}_{\alpha}|\rho},\label{eq:resolveddensitycalculation}\\
\rho_0 \underline{u}_i&=&\sum_{\alpha=0}^{8}\underline{\overline{f}}_{\alpha} e_{\alpha i}+\frac{1}{2}\underline{\rho}\thinspace\thinspace\underline{a^{'}_i}=\braket{\underline{\overline{f}}_{\alpha}|e_{\alpha i}}+\frac{1}{2}\underline{\rho}\thinspace\thinspace\underline{a^{'}_i}. \qquad i \in {x,y} \label{eq:resolvedvelocitycalculation}
\end{eqnarray}
This provides a minimal lattice kinetic equation for incorporating turbulence models, where the unresolved turbulent motion
are inertial frame invariant with respect to the resolved fluid motion. Here, we clarify the meaning of this statement as follows.
Unlike other areas in fluid mechanics, where models have been developed starting from continuous kinetic theory, its role for fluid turbulence has been more limited. This is mainly due to the fact that kinetic theory generally considers distinct scale separation of physical processes. On the other hand, turbulence is a flow phenomenon intrinsically containing a continuous spectrum of scales with no
scale separation. As such, therefore, turbulence modeling developments have to rely much on phenomenology whose mathematical forms
are then constrained by invariance principles (e.g. material frame indifference and inertial frame invariance mentioned earlier in the introduction) and realizability considerations~\cite{fureby97,durbin10}. Thus, except for some early models such as those based on mixing length concepts and derivation of some recent phenomenological models (e.g.~\cite{degond02}) based on kinetic theory, turbulence modeling developments are generally based on macroscopic models. The ultimate goal of our central moment approach for the filtered kinetic equation discussed above, is, then, to simulate resolved turbulent fields which are inertial frame independent, when an appropriate macroscopic turbulence model for the unresolve field is used in the forcing term.

\section{\label{sec:summary}Summary and Conclusions}
A discrete lattice kinetic model for the continuous Boltzmann equation, including forcing, based on central moments is derived. The collision operator as well as the source term of this lattice Boltzmann equation is constructed by matching the corresponding
continuous and discrete central moments successively at various orders. The local attractor of the collision operator is
constructed to satisfy the factorization property of the Maxwellian during relaxation process. An exact hierarchical identity
of the central moment of sources, that incorporates non-equilibrium effects, is maintained at the discrete level. The resulting
approach provides Galilean invariant hydrodynamic fields in the presence of any external or self-consistent internal forces in a discrete kinetic framework. It is further extended to incorporate reduced compressibility effects for better representation of incompressible flow, a limiting case. An important physical characteristic of turbulent flows is that it is inertial frame independent for all or any subset of scales. In consequence, for general applicability, all turbulence models and their simulation approaches, should satisfy this requirement. A statistical averaged/filtered lattice kinetic equation based on central moments that maintains Galilean invariant representation of
unresolved fluid motion with respect to the resolved fields of turbulent flow is thus constructed. The formalism presented
here can extended to other lattice velocity sets and in three-dimensions as well as to other physical problems such as complex fluids.

In this regard, we make the following remark on the development of more efficient schemes for the former aspect. Symmetry and finiteness of the standard lattice sets lead to degeneracies of higher order moments in terms those at lower orders that can result in frame dependent contributions to viscous stresses. This necessitates considerations of extended lattice sets. In this case, it is proposed that the \emph{central moments} relaxation (as well as forcing) be considered \emph{only} up to those higher order moments that have bearing on the physics of \emph{hydrodynamics}, such as stress tensors and heat flux vectors. In turn, this would impose Galilean invariance of the macroscopic description of the fluid motion. The relaxation of the rest of the higher moments (including forcing) related to the fast \emph{kinetic or ghost modes} can be considered in terms of the standard or \emph{raw moments}. Here, the form of the hierarchical identity for the sources derived in this paper for those higher (kinetic) moments would be the same with the simple replacement of the central moment terms by the corresponding raw moments. It is envisaged that such mixed central/raw moment approach would exhibit interesting mathematical structures for the resulting collision operator and the sources, as well as being computationally more effective. This strategy is currently under investigation.


\end{document}